\documentclass[12pt,a4paper]{article}

\usepackage{amsmath,amsfonts,latexsym,amssymb,amsthm}
\usepackage{verbatim,amsthm,curves,graphics}
\usepackage{mathrsfs}

\begin{document}

\theoremstyle{definition}
\newtheorem{defn}{Definition}[section]
\newtheorem{remark}{Remark}[section]
\newtheorem{thm}{Theorem}[section]
\newtheorem{lem}{Lemma}[section]
\newtheorem{prop}{Proposition}[section]
\renewcommand{\thesection}{\Roman{section}}
\newcommand{\tr}{\operatorname{tr}}
\newcommand{\vol}{{\rm vol}}
\newcommand{\g}{{g}}
\renewcommand{\i}{{\rm i}}
\newcommand{\supp}{\operatorname{supp}}
\newcommand{\sign}{\operatorname{sign}}
\newcommand{\id}{\operatorname{id}}
\newcommand{\cO}{\mathcal{O}} 
\newcommand{\cC}{\mathscr{C}} 
\newcommand{\cE}{\mathscr{E}} 
\newcommand{\cM}{\mathscr{M}} 
\newcommand{\cD}{\mathscr{D}} 
\newcommand{\cA}{\frak{A}} 
\newcommand{\fF}{\frak{F}} 
\newcommand{\cW}{\frak{W}} 
\newcommand{\cK}{\mathscr{K}} 
\newcommand{\cH}{\mathscr{H}} 
\newcommand{\cV}{\mathscr{V}} 
\newcommand{\cI}{\mathscr{I}} 
\newcommand{\cF}{\mathscr{F}} 
\newcommand{\cN}{\mathscr{N}} 
\newcommand{\cX}{\mathscr{X}} 
\newcommand{\cY}{\mathscr{Y}} 
\newcommand{\cR}{\mathscr{R}} 
\newcommand{\mn}{\Bbb{N}} 
\newcommand{\ben}{\begin{equation}} 
\newcommand{\een}{\end{equation}} 

\newcommand{\mc}{\Bbb{C}} 

\newcommand{\mi}{\Bbb{I}} 

\newcommand{\mj}{\Bbb{J}} 

\newcommand{\mr}{\Bbb{R}} 

\newcommand{\mslash}{/\!\!\!}             
\newcommand{\slom}{/\!\!\!G}         
\newcommand{\dirac}{/\!\!\!\nabla}        
\newcommand{\myid}{\leavevmode\hbox{\rm\small1\kern-3.8pt\normalsize1}}
\newcommand{\rso}{|\!|\!|}
\newcommand{\lso}{|\!|\!|}

\title{Nuclearity, Local Quasiequivalence and Split Property for Dirac
Quantum Fields in Curved Spacetime}
\author{Claudio D'Antoni\\
        \it{Dipartimento di Mathematica, Universit\`a 
            di Roma ``Tor Vergata'',}\\ 
        \it{Via della Ricerca Scientifica,}\\
        \it{I-00133 Roma, Italy}
\and
        Stefan Hollands\\    		
        \it{Department of Physics, Enrico Fermi Institute,}\\
	\it{University of Chicago, 
            5640 Ellis Ave.,} \\ 
        \it{Chicago, IL 60367, U.S.A.}
}
\date{\today}

\maketitle

\begin{abstract}
We show that a free Dirac quantum field on a globally hyperbolic
spacetime has the following structural properties: (a) any two 
quasifree Hadamard states on the algebra of free Dirac fields are 
locally quasiequivalent; (b) the split-property 
holds in the representation of any
quasifree Hadamard state; (c) if the underlying spacetime is static, 
then the  nuclearity condition is satisfied, that is, the 
free energy associated with a finitely extended subsystem (``box'')
has a linear dependence on the volume of the box and goes like 
$\propto T^{s+1}$ for large temperatures $T$, where $s+1$ is the number of
dimensions of the spacetime.
\end{abstract}

\section{Introduction}

In the  {\it algebraic framework} of quantum field theory 
\cite{haag}, one takes the point of view that quantum field theoretical models 
should ultimately be described purely in terms
of their associated algebras of local observables. The basic mathematical
structure in this approach is an  assignment $\cO \to \cA(\cO)$ of 
finite regions in spacetime with $C^*$-algebras $\cA(\cO)$, containing the  
observables in the theory that are localized in $\cO$. 
This assignment is expected to have some general, model-independent  
structural features; namely, if $\cO_1 \subset \cO_2$, then 
$\cA(\cO_1) \subset \cA(\cO_2)$ and if $\cO_1$ and $\cO_2$ 
are spacelike separated, then the elements in $\cA(\cO_1)$ 
should commute with those in $\cA(\cO_2)$. 
The above properties express in an 
abstract way  the general features of locality and local 
commutativity of a quantum field theory.  

A quantum state in the algebraic framework is a  
positive normalized  linear functional $\omega:\cA \to \mc$. It is a 
basic fact from the theory of $C^*$-algebras that every
state determines a representation $\pi$, called the 
GNS-representation,  of $\cA$ on 
some Hilbert space $\cF$ containing a distinguished vector 
$|\Omega\rangle$ such that $\omega(A) = \langle \Omega |
\pi(A) \Omega\rangle$ for all $A \in \cA$. 
If the spacetime under consideration 
is Minkowski space, one furthermore postulates
the existence of a preferred vacuum state, $\omega_0$, whose 
GNS-representation carries a positive energy representation 
of the translation symmetry group of Minkowski space. 

The strength of the algebraic approach is 
that a number of structural features known from 
quantum field theory can be seen to be a consequence of 
general and model-independent axioms, 
see Haag's book \cite{haag} for an overview. 
However, these axioms do not yet entail such basic features as the 
existence of particles or a reasonable thermodynamic behavior. 
What is missing is a suitable mathematical formulation of the 
idea that excitations of the vacuum localized in 
finite regions and with finite energy occupy ``finite volumes
in phase space''. Haag and Swieca \cite{hs} suggested to incorporate 
this idea into the general framework of algebraic quantum field 
theory in Minkowski
space by demanding that a reasonable theory should satisfy the following 
{\it compactness criterion}: Let $\cO_r$ be a double cone 
in Minkowski space whose base is a ball of radius $r$
and let $\pi_0$ be the GNS-representation of the vacuum 
state on the Hilbert space $\cF_0$, with vacuum vector 
$|\Omega_0\rangle$. 
Then 
the subsets of $\cF_0$ given by 
$$
\{ P_E \pi_0(A)|\Omega_0\rangle \in \cF_0 \mid A \in \cA(\cO_r), \|A\| \le 1 \}
$$
are compact\footnote{
Recall that a subset of a Banach space is called compact if every 
bounded sequence in this set contains a weakly convergent subsequence.}
for all $E$ and $r$, where $P_E$ is the spectral projector 
of the Hamiltonian, $H$, to the interval $[0, E]$. 

With the same idea in mind, 
Buchholz and Wichmann \cite{buchholz&wichmann} proposed a 
stronger requirement on the 
size of the phase space volumes corresponding to excitations 
localized in finite regions in Minkowski space and with finite energy. 
Instead of a sharp cut off in energy (represented 
above by the projector $P_E$) they use an exponential damping and 
replace the above compactness criterion by the following 
{\it nuclearity requirement:} The maps $\Theta_\beta$ defined by 
\begin{equation}
\label{eq1}
\cA(\cO_r) \owns A \to e^{-\beta H} \pi_0(A) |\Omega_0\rangle = 
\Theta_\beta(A) 
\in \cF_0
\end{equation}
are nuclear for all $\beta > 0$, with {\it nuclearity index} bounded by 
$\nu_{r, \beta} \le \exp[c r^s/\beta^{n}]$, where 
$c, n$ are positive constants, $s$ is the number of spatial 
dimensions and $r$ the radius of the base of 
$\cO_r$ (the definition of a nuclear map and 
its nuclearity index is given Sec. \ref{sec2.2}). 
The physical interpretation of
$\nu_{r, \beta}$ is that of the {\it partition function} 
of the finitely extended subsystem located in $\cO_r$ at 
temperature $T = \beta^{-1}$. The 
dependence of $\nu_{r, \beta}$ on $r$ then expresses that the free energy, 
$F_{r, \beta}= -\beta^{-1} \ln \nu_{r, \beta}$, of 
the finitely extended system depends at most linearly on the volume of 
the box. The dependence on $\beta$ gives the relation between energy 
and temperature in the high temperature region. It has been shown 
to be sufficient to guarantee normal thermodynamic behavior. 

The above nuclearity condition, as well as different variants thereof (see for example \cite{bp}), 
have been verified for free scalar fields and, more generally, for suitable infinite multiplets
of these fields in Minkowski space \cite{bp}. A suitably modified nuclearity 
condition can also be formulated in curved spacetimes with a time translation 
symmetry, for which there exists a state $\omega_0$ carrying a 
positive energy 
representation of the time translation group (so that $H$ now denotes the 
generator of this symmetry). That the nuclearity property 
still holds under these more general circumstances was established 
by Verch \cite{v} for the case of a free
scalar field in the representation of the natural ground state 
on an ultrastatic spacetime.\footnote{Actually he verified a  
version of nuclearity condition, which is stronger than the one given above.} 

It is expected that the nuclearity property should hold also for free 
fields with higher spin, 
but a discussion of this seems to be missing 
from the literature so far, even in the case of Minkowski space. 
One purpose of the present paper is to fill this gap for the 
spin-$\frac{1}{2}$ case. Namely, we will prove that the 
nuclearity condition holds for the free Dirac field on a static spacetime
in the GNS-representation of the ground state, and that the nuclearity index, 
$\nu_{\cC, \beta}$, associated with a double cone $\cO_\cC$ with 
compact base $\cC$, satisfies the estimate
\begin{equation}
\label{eq2}
\nu_{\cC, \beta} 
\le \exp\left[\left(\frac{\beta_0}{\beta}\right)^s e^{-m_0 \beta/2}\right]. 
\end{equation}
Here, $\beta_0$ is a constant depending on
on the spatial geometry of the spacetime under consideration 
and $m_0$ is a constant proportional to the mass of the field. 
This estimate implies that the free energy $F_{\cC, \beta}
= - \beta^{-1} \ln \nu_{\cC, \beta}$ 
associated with the subsystem localized in $\cO_\cC$ 
goes at most like $T^{s+1}$ for high temperatures $T$. 
We also show that if the spacetime metric is 
rescaled by $g_{ab} \to \lambda^{2} g_{ab}$, then the 
nuclearity index for the rescaled spacetime is estimated by 
the expression on the right side of \eqref{eq2}, with 
$\beta_0$ replaced by $\lambda \beta_0$. This means 
that the free energy $F_{\cC, \beta}$ has at most a linear
dependence on $\lambda^s$ under rescalings of the metric, and 
therefore at most a linear dependence on 
the volume of the box.\footnote{Note that, in Minkowski space, 
the transformation $g_{ab} \to \lambda^2 g_{ab}$ is 
equivalent, via the conformal isometry $x \to \lambda x$, to a 
rescaling of the box itself, $\cO_r \to \cO_{\lambda r}$.} 

The proof of the above statements does not readily follow from 
the arguments developed for scalar fields, because the underlying 
combinatorics in both cases are quite different, 
due to the different statistics of bosonic and fermionic fields.
Also, the above estimate for $\nu_{\cC, \beta}$ cannot readily be obtained by the same methods as  
in Minkowski space. Instead, we obtain it using some results and methods 
from the theory of pseudo differential operators.
We mention as an aside that our proof of the nuclearity property 
also shows that the map \eqref{eq1} is actually even $p$-nuclear 
for all $p > 0$.

A quantum field model on a curved spacetime 
$M$ is said to satisfy the so-called {\it split-property} 
(with respect to some state, $\omega$) 
if the following holds: Let $\cO_1$ and $\cO_2$ be two concentric open
double cones with bases $\cC_1$ and $\cC_2$ (contained in some Cauchy-surface $\Sigma$) 
such that ${\rm clo}(\cC_1) \subset \cC_2$.
Furthermore, let $\pi_\omega(\cA(\cO_1))''$ and $\pi_\omega(\cA(\cO_2))''$ be the 
weak closures of the corresponding local algebras in the GNS-representation
of the state $\omega$. Then there exists a type I factor $\cN$ such that
$$
\pi_\omega(\cA(\cO_1))'' \subset \cN \subset \pi_\omega(\cA(\cO_2))''.
$$
It is known \cite{bdf} that the split-property automatically holds 
for the ground state $\omega_0$ on an ultrastatic spacetime if the 
nuclearity condition is satisfied.
Hence our estimate \eqref{eq2} immediately gives us 
that the split-property holds for free Dirac fields in static 
spacetimes in the GNS-representation of the ground state.
(For the special case of a Dirac field in Minkowski spacetime this has 
previously been shown by Summers \cite{sum}.) 
Combining this with a deformation argument due to Verch \cite{v}, we 
moreover show in Sec.~\ref{sec4} that the split-property holds in fact 
on an arbitrary globally hyperbolic spacetime
in the GNS-representation of any quasifree Hadamard 
state (an explanation of this term will be given below). 
Our proof of this partly relies on 
the fact that any two quasifree Hadamard states $\omega_1$ and $\omega_2$
for the Dirac field are {\it locally quasiequivalent}, meaning that the density matrix states in 
the GNS Hilbert spaces of $\omega_1$ and $\omega_2$ define the same sets of states of
$\cA(\cO)$ for any relatively compact subset $\cO$ of $M$. 
The analogue of this result for scalar fields had 
previously been obtained by Verch \cite{ver2}, 
but a proof for Dirac case seems to be 
missing from the literature so far.\footnote{For scalar fields, 
it has recently been proven \cite{js} that local  
quasiequivalence still holds under the much weaker assumption that 
the states in question are not Hadamard but only 
``adiabatic of order $N$'', for some suitable number $N$. One 
would expect that an analogous result holds also for Dirac fields.
A first result in this direction on Robertson-Walker spacetimes
has been obtained by \cite{h}.} 
It is therefore provided in Sec.~\ref{sec3}. In the course of this 
proof we also show that the local v.~Neumann algebras 
$\pi_\omega(\cA(\cO_\cC))''$ associated with a double cone 
$\cO_\cC$ (with regular base) is a {\it factor} 
for any quasifree state $\omega$.
 
Because all quasifree Hadamard states are locally quasiequivalent, 
the v. Neumann algebras $\pi_\omega(\cA(\cO))''$ are algebraically
isomorphic for different choices of the quasifree state $\omega$. They 
may therefore regarded as different realizations of some abstract
factorial v. Neumann algebra $\cR(\cO)$.
The collection of these factors can be shown to have the following
{\it general covariance property} (for a detailed discussion 
of this property, see \cite{ver3,hw}): 
Let $\chi: N \to M$ be a causality-preserving isometric 
embedding of a spacetime 
$(N, h_{ab})$ into a larger spacetime $(M, g_{ab})$, so that 
$\chi^* g_{ab}= h_{ab}$, and assume that $\chi$ induces a homomorphism of the 
corresponding spin-structures.
Then there exists a homomorphism $\alpha_\chi$ such that 
$\alpha_\chi(\cR(\cO))=\cR(\chi(\cO))$ for all relatively compact 
$\cO \subset N$.\footnote{As pointed out to us by the referee,
in order that $\alpha_\chi$ be independent of $\cO$, one must 
actually require that the isometric embedding $\chi$ is extendible 
to some $N'$ containing $N$ as a subset with compact closure.}
Moreover, if $\chi_1 \circ \chi_2$ is 
the composition of two isometric embeddings, then 
it holds that $\alpha_{\chi_1\circ\chi_2} = 
\alpha_{\chi_1} \circ \alpha_{\chi_2}$. 
In the case when $\chi$ is a diffeomorphism, 
the map $\alpha_\chi$ implements the transformation 
of local quantum fields under $\chi$. 
Now if $\cO_1 \subset \cO_2 \subset N$ are open regions as 
in the discussion of the split-property, and if $\cN$ is 
a type I factor corresponding to this inclusion, then 
$\alpha_\chi(\cN)$ is a type I factor
corresponding to 
the inclusion $\chi(\cO_1) \subset \chi(\cO_2) \subset M$. 
In this sense, the type I factor $\cN$ in 
the split-property `transforms in a generally covariant way'. 
 
The plan and main results of this paper may now be summarized: In 
Section~\ref{sec2.1}, we 
review the quantization of a free Dirac field on a curved spacetime 
from the algebraic point of view and 
recall the notion of quasifree and Hadamard states. 
In Section~\ref{sec3.1} we show 
that any two quasifree Hadamard states are locally quasiequivalent
(Thm.~\ref{lqe}), and that the local v.~Neumann algebras
associated with a quasifree state are factors of type 
$III_1$ (Thm.~\ref{lf}).  Section~\ref{sec3} contains our results concerning 
the nuclearity property for Dirac fields in static spacetimes (Thm.~\ref{main} and Prop.~\ref{prop}).  
The split-property for Dirac fields in arbitrary globally hyperbolic 
spacetimes is established in the last section. 
The definition of nuclear maps and some related functional
analytic concepts is given in the Appendix.

\medskip

{\it Notations and conventions:}
Throughout, $(M, \g_{ab})$ denotes an oriented, time-oriented  globally hyperbolic 
spacetime of dimension $s+1$. The metric volume element compatible 
with the orientation of $M$ is denoted by ${\bf \epsilon}_{a_0
a_1 \dots a_s}$. We denote by $J^\pm(\cO)$ 
the causal future respectively past of a region $\cO \subset M$, i.e., 
the set of all points that can be reached by a future
respectively past directed causal curve starting in $\cO$. 
$D(\cO)$ denotes the domain of dependence of $\cO$, 
defined as the set of all points $x$, such any 
future and past inextendible causal curve through $x$ intersects  
$\cO$. Our signature convention is $+---\dots$. 

\section{Review of classical and quantum Dirac fields in curved spacetimes}
\label{sec2.1}

\subsection{Classical fields}

We begin by reviewing briefly 
the the theory of a classical, linear Dirac field on a curved 
spacetime (for a more detailed discussion see for example \cite{dim1,sv1}). 
In order to define spinors on a 
curved spacetime, it is necessary to first introduce the 
notion of a spin-structure. 
Let $FM$ be the bundle of all oriented, 
time-oriented orthonormal frames over $(M, g_{ab})$. 
$FM$ has the structure of a principal fibre bundle 
with base manifold $M$ and structure group  
${\rm SO}^\uparrow(s, 1)$, where the arrow indicates that the 
transformations preserve the time-orientation. The universal
2--1 covering group of ${\rm SO}^\uparrow(s, 1)$ is the group 
${\rm Spin}^\uparrow(s, 1)$. Elements $g$ of this group can be
represented by complex $2^{[(s+1)/2]}$--dimensional\footnote{
$[\dots]$ denotes the integer
part of a number.} square matrices 
satisfying 
\begin{equation}
\label{mgamma}
g\gamma^\mu g^{-1} = \Lambda(g)^\mu_\nu \gamma^\nu. 
\end{equation}
Here, the $\gamma^\mu$ denote a set of $2^{[(s+1)/2]}$--dimensional 
gamma-matrices for $(s+1)$--dimensional Minkowski spacetime\footnote{
We assume that the gamma matrices in \eqref{mgamma} are taken in the 
Majorana representation, which requires that $s+1 = 
3,4,9,10 \,{\rm mod} \,8$.}  
satisfying
\begin{equation}
\gamma^\mu \gamma^\nu + \gamma^\nu \gamma^\mu = 2 \eta^{\mu\nu}, 
\end{equation} 
and $\Lambda$ denotes the covering homomorphism.
A spin structure on $(M, g_{ab})$ is by definition a principal 
fibre bundle $SM$ with structure group ${\rm Spin}^\uparrow(s, 1)$ 
and base manifold $M$, together with a 
bundle homomorphism $SM \to FM$ compatible with the covering
homomorphism $\Lambda$. Whether a given spacetime admits a spin 
structure or not depends only on the topology of $M$. In the following
we assume that $M$ is such that a spin structure exists. Given a spin structure, 
one defines the ``spinor bundle'', $DM$, as 
the $2^{[(s+1)/2]}$--dimensional
complex vector bundle over 
$M$ which is associated to $SM$ 
via the representation of ${\rm Spin}^\uparrow(s, 1)$
on $\mc^{2^{[(s+1)/2]}}$ defined by \eqref{mgamma}.

In order to write down the 
Dirac equation in $M$, one needs an 
analogue of Gamma-matrices as well as a suitable 
covariant derivative operator for spinor fields in curved spacetime. 
Gamma matrices $\gamma^a$ in curved spacetime are 
the fiber-wise linear maps in $DM$ satisfying
$\gamma_a \gamma_b + \gamma_b \gamma_a = 2\g_{ab}$.
They are defined by requiring that their components with respect to an 
appropriate local frame are equal to those of the corresponding
gamma matrices \eqref{mgamma} in Minkowski space.
Derivatives of spinor fields in 
curved spacetime can naturally be defined in terms of the 
so-called ``spin-connection'', $\nabla_a$, which is 
the uniquely determined covariant derivative operator 
satisfying $\nabla_a \gamma_b = 0$. The Dirac equation
for a spinor field $u$ in curved spacetime then reads
\begin{eqnarray}
\label{dirac}
(i\gamma^a\nabla_a - m) u = 0, 
\quad m \ge 0. 
\end{eqnarray}
It is known \cite{dim1} that there exist unique
advanced and retarded fundamental
solutions ${\cal S}^{\rm adv}$ and ${\cal S}^{\rm ret}$ to this equation, 
i.e., continuous linear maps from $\cD(DM)$ 
(the space of smooth spinor fields with compact support) to $\cE(DM)$ 
(the space of smooth spinor fields), 
for which there holds
\begin{eqnarray*}
(i\gamma^a\nabla_a - m){\cal S}^{\rm adv} = 
{\cal S}^{\rm adv}(i\gamma^a\nabla_a - m) = \myid, 
\end{eqnarray*}
(with the same equation holding for ${\cal S}^{\rm ret}$)
and for which $\supp({\cal S}^{\rm adv} u) \subset J^+(\supp(u))$
and $\supp({\cal S}^{\rm ret}u) \subset J^-(\supp(u))$. 
The causal propagator, $\cal S$, is the distributional 
bisolution defined by 
${\cal S} = {\cal S}^{\rm adv} - {\cal S}^{\rm ret}$.  

The notion of the charge-conjugate and the Dirac-conjugate 
of a spinor field $u$ in Minkowski spacetime
can be generalized in a natural and invariant way to spinor
fields on a curved spacetime via an appropriately 
chosen trivialization of $DM$ (the charge conjugation is defined only 
for $s+1 = 3, 4, 9, 10 \,{\rm mod}\,8$). 
The charge conjugate of a spinor field $u$ 
is written as $Cu$, where $C$ is a certain anti-linear, 
fiber-preserving, involutive ($C^2 = \myid$) map from  
$DM$ to itself. If $u$ is a solution to the Dirac equation, 
then so is $Cu$. Spinor fields which are invariant under the charge 
conjugation map are called ``Majorana spinors''. 
The Dirac-conjugate of a spinor field, $\bar u$, lives in the 
dual, $D^*M$, of the spinor bundle. 
If $u$ is  a solution to the Dirac equation~\eqref{dirac}, then 
$\bar u$ is a solution to $\bar u(-i\gamma^a\nabla_a - m) = 0$,
and vice versa. 

\subsection{Quantum fields}

The theory of a quantized Dirac field can be formulated in various 
ways. We here present the theory from the algebraic point of 
view, using Araki's selfdual framework \cite{dim}. For simplicity, 
we will restrict attention from now on to Majorana fields (which 
means that $s+1 = 3, 4, 9, 10 \,{\rm mod}\,8$), but 
our results can easily be generalized to charged Dirac fields. 

The algebra $\fF$ of Dirac-Majorana quantum fields is by definition the
uniquely determined, unital $C^*$-algebra generated by elements
of the form $\Psi(u)$ (smeared Dirac-Majorana quantum fields), where 
$u$ is a compactly supported, smooth spinor field. They are 
subject to the following relations: 

\medskip
\noindent
{\bf Linearity:} The assignment $\cD(DM) \owns u \to \psi(u) \in \fF$ is complex linear.\\ 
{\bf Dirac Equation:} There holds $\Psi((i\gamma^a\nabla_a - m)u) = 0$. \\
{\bf Canonical Anticommutation Relations:} $\{ \Psi(u_1), \Psi(u_2) \} = 
i{\cal S}(C \bar u_1, u_2)\myid$, where $\cal S$ is viewed as a
bilinear form on $\cD(D^*M) \times \cD(DM)$. \\
{\bf Hermiticity:} $\Psi(u)^* = \Psi(Cu)$. 

\medskip
\noindent
The local field algebras $\fF(\cO)$ corresponding to some region $\cO$ in spacetime, 
are by definition the $C^*$-subalgebras of $\fF$ generated by all elements of the form $\Psi(u)$, 
where $\supp(u) \subset \cO$. 

In the following we will often make use of an alternative description of the algebra $\fF$ 
in terms of the Cauchy data of the field operators on some Cauchy surface 
$\Sigma$ (``sharp time fields''), as we now briefly explain. It can be seen that 
\begin{eqnarray}
\label{ip}
(k_1 | k_2) = 
\int_\Sigma \bar k_1 \gamma^a k_2 \, {\bf \epsilon}_{ab_1 \dots b_s}
\end{eqnarray} 
defines a positive, non-degenerate Hermitian inner 
product on the space of all compactly supported smooth spinor fields on 
$\Sigma$. The completion of this space with respect to 
this inner product is a Hilbert space which we will denote as $\cK = 
L^2(DM \restriction \Sigma)$. 
The charge conjugation map $C$ can be seen to be 
antiunitary with respect to that inner product, that is, 
$(Ck_1| Ck_2) = (k_2 | k_1)$ for all $k_1, k_2 \in \cK$.
One can show that $\fF$ is $^*$-isomorphic to the $C^*$-algebra generated
by elements of the form $\psi(k)$ with $k \in \cK$, 
subject to the relations $\psi(Ck) = \psi(k)^*$ and 
$\{\psi(k_1), \psi(k_2)\} = (Ck_1 | k_2)\myid$.
The isomorphism is explicitly given by 
\ben
\label{sharptime}
\Psi(u) \to \psi(k), \quad k = {\cal S} u\restriction \Sigma, 
\een
for spinor fields $u \in \cD(DM)$. 
The local field algebras $\fF(\cO)$, where $\cO$ is a region of the form
$D(\cC)$ ($\cC$ an open subset of $\Sigma$) correspond, 
under this isomorphism, to the algebras generated by elements of the form 
$\psi(k)$, where $k \in L^2(DM \restriction \cC)$. 

The algebra $\fF$ has the following structural property \cite{dim}: 
Any unitary operator $U$ on $\cK$ which commutes with $C$ induces a unique automorphism $\alpha$ on 
$\fF$ satisfying $\alpha(\psi(k)) = \psi(Uk)$ for all $k \in \cK$. $\alpha$ is called the 
``Boguliubov automorphism'' associated with $U$. 

\medskip

A state in the algebraic framework is by definition a linear functional
$\omega: \fF \to \mc$, which is normalized so that $\omega(\myid) = 1$
and positive in the sense that $\omega(F^* F) \ge 0$ for all $F \in \fF$. 
The algebraic notion of states is related to the usual Hilbert space 
notion of states
by the GNS-theorem. This says that for any algebraic state, there is 
a representation $\pi_\omega$ of $\fF$ as bounded, linear operators 
on a Hilbert space $\cF_\omega$ containing a distinguished vector, 
$|\Omega_\omega \rangle$, such that $\omega(F) = \langle \Omega_\omega |
\pi_\omega (F) \Omega_\omega \rangle$ for all $F \in \fF$. The 
two-point function, ${\cal S}_\omega$, of a 
state $\omega$ on $\fF$ is the bidistribution on $M \times M$ defined
by 
$${\cal S}_\omega(u_1, u_2) = \omega(\Psi(u_1)\Psi(u_2))$$ 
for all smooth spinor fields 
$u_1$ and $u_2$ over $M$. 

In this paper, we will restrict attention to a particular 
class of states on $\fF$, namely the so-called ``quasifree 
states''. A quasifree state is one for which\footnote{An equivalent 
way of saying that a state is quasifree is to demand that it has  
a vanishing one-point function, $\omega(\psi(u)) = 0$, and 
vanishing truncated $n$-point functions for $n > 2$. 
}
\begin{eqnarray}
\label{qf}
\omega(\Psi(u_1) \dots \Psi(u_{n})) = 
\begin{cases}
\sum_{p} {\rm sign}(p) \prod_{(i,j)\in p} \omega(\Psi(u_i)\Psi(u_j)), 
&\text{for $n$ even,}\\
0 &\text{for $n$ odd,}
\end{cases}
\end{eqnarray}
where the sum is over all partitions $p$ of $\{1, \dots, n\}$ 
into pairs $(i,j)$ with $i < j$, and where 
${\rm sign}(p)$ is the signature of the permutation  
${\scriptsize \left(
\begin{array}{ccccc}
n & n-1 & \dots & 2 & 1 \\
j_1 & i_1 & \dots & j_{n/2} & i_{n/2}
\end{array}
\right)}$. The above formula implies in particular that 
a quasifree state is entirely specified by its two-point function.

It is not difficult to see that the two-point function of a 
quasifree state can always be written in the form
\begin{equation}
\label{fst}
{\cal S}_\omega(u_1, u_2) = \Big(C\rho_\Sigma {\cal S}u_1 \Big|
P \rho_\Sigma {\cal S} u_2 \Big) \quad \text{for all
$u_1, u_2 \in \cD(DM)$,} 
\end{equation}
where $\rho_\Sigma$ means the operation of restricting a spinor field
over $M$ to the Cauchy surface $\Sigma$, $\cal S$ is the 
causal propagator and where $P$ is a bounded operator on 
$\cK$ satisfying $0 \le P \le \myid$ and $CPC = \myid - P$. In terms of 
the sharp time fields $\psi(k)$ (see eq.~\eqref{sharptime}), 
the two-point function is simply
\ben
\label{twopoint}
\omega(\psi(k_1) \psi(k_2)) = (Ck_1|Pk_2).
\een 
Conversely, given any operator $P$ on $\cK$ with the above 
properties, eq.~\eqref{twopoint} defines a quasifree state on $\fF$. 
If $P$ is in addition a projection operator, 
$P^2 = P$, then the corresponding quasifree state is called a 
``Fock state''. A quasifree state is pure if and only if it is a 
Fock state. 
The GNS representation of $\fF$ associated with a quasifree state is 
described as follows: $\cF_\omega$ is the 
antisymmetric Fock-space over the ``1-particle Hilbert space''
\begin{equation}
\label{Hdef}
\cH = {\rm clo} \{ P^{1/2} k \mid k \in C^\infty_0(DM \restriction \Sigma) \},
\end{equation} 
that is, 
\begin{equation}
\label{fock}
\cF_\omega = \mc \oplus \big(\oplus_{n \ge 1} (\wedge^n \cH) \big),
\end{equation}
where $\wedge^n \cH$ is the $n$-th antisymmetrized tensor 
power of $\cH$. The Hilbert space $\cF_\omega$ is thus the closure of 
the linear span of vectors of the form
\ben
|\Phi\rangle = q_1 \wedge q_2 \wedge \cdots \wedge q_n, \quad q_i \in \cH, 
\een
and the ``vacuum vector'', $|\Omega_\omega \rangle$, corresponding to the 
first component $\mc$ in the direct sum~\eqref{fock}. The scalar product 
in $\cF_\omega$ is that naturally induced from the scalar product $(\, | \,)$ 
in $\cH$. The GNS-representation is
\begin{eqnarray}
\label{repdef}
\pi_\omega(\psi(k)) = a(P^{1/2}k)^* + a(P^{1/2}Ck) \quad \forall k \in \cK,
\end{eqnarray}
where $a(p)^*$ and $a(p)$ with $p \in \cH$ are the creation respectively annihilation operators defined by\footnote{Note that
$p \to a(p)^*$ is linear in $p$ and that $p \to a(p)$ is antilinear in $p$.} 
\begin{eqnarray}
\label{adef}
a(p)^* (q_1 \wedge q_2 \wedge \cdots \wedge q_n) &=& p \wedge q_1 \wedge q_2 \wedge \dots \wedge q_n, \nonumber\\
a(p) (q_1 \wedge q_2 \wedge \cdots \wedge q_n) 
&=& \sum_{r=1}^n (-1)^{r+1} (p|q_r)  q_1 \wedge \dots \widehat{q_r} \wedge \dots q_n, \quad q_i, p \in \cH, 
\end{eqnarray}
as well as the action 
\ben
a(p)^*|\Omega_\omega\rangle = p, \quad a(p) |\Omega_\omega\rangle = 0
\een
on the vacuum vector. They satisfy the usual anti-commutation relations
\ben
\{ a(p_1), a(p_2)^* \} = (p_1 | p_2). 
\een
Since the generators $\psi(k)$ of the algebra $\fF$ satisfy anticommutation 
relations, one has a slighly different notion of locality as
compared to the case of Bose fields. In order to describe this 
notion of locality, it is convenient to introduce the grading 
automorphism $\gamma$ on $\fF$,  which is defined 
by $\gamma(\psi(k)) = -\psi(k)$ in terms of the generators of $\fF$. We have 
that $\gamma(F) = \pm F$ for $F \in \fF_\pm$, where $\fF_+$ is generated by
even products of $\psi$, and where $\fF_-$ is generated by odd products of $\psi$. 
Each $F$ can be uniquely decomposed into parts from $\fF_+$ and $\fF_-$ as 
$F = F_+ + F_-$, where $F_\pm = \frac{1}{2}(F \pm \gamma(F))$. Given this 
decomposition, one can introduce the graded commutator by 
\ben
\label{gracom}
[F_1, F_2]_\gamma = 
\begin{cases}
F_1 F_2 + F_2 F_1 & \text{if $F_1, F_2 \in \fF_-$},\\
F_1 F_2 - F_2 F_1 & \text{otherwise,}
\end{cases}
\een
where we are assuming without loss of generality that $F_i \in \fF_\pm$. 
It is a consequence of the canonical anticommutation relations, together
with locality properties of the anti-commutator function $\mathcal S$ that
\ben
\label{gradloc}
[F_1, F_2]_\gamma = 0 \quad \text{if $F_1 \in \fF(\cO_1),F_2 \in \fF(\cO_2)$,} 
\een
and if $\cO_1$ and $\cO_2$ are spacelike separated.
The even part $\fF_+(\cO)$, but not the odd part, is a subalgebra of 
$\fF(\cO)$. Since the graded commutator is just the ordinary 
commutator for elements in the even part, it follows that
\begin{equation}
[F_1, F_2] = 0 \quad \text{if $F_1 \in \fF_+(\cO_1),F_2 \in \fF_+(\cO_2)$,}
\end{equation}
whenever $\cO_1$ and $\cO_2$ are spacelike separated.
The algebra $\fF_+(\cO)$ therefore satisfies bosonic commutation relations 
appropriate for true observables in the theory. We will refer to it as the 
``algebras of observables'', and we will sometimes use the notation $\cA(\cO)$ for this
algebra.

Actually, one can get rid of the graded commutator in eq.~\eqref{gracom} in favor
of the ordinary commutator via a standard construction often referred to as ``twist'', and 
this leads to a slightly more subtle notion of locality, called ``twisted locality'', 
for the algebra of fields, $\fF(\cO)$. 
We now briefly recall this concept, since it will play a role below. 
Any quasifree state $\omega$ is invariant under the 
automorphism $ \gamma $,
that is, $\omega(\gamma(F)) = \omega(F)$ for all $F \in \fF$, because such a state 
by definition vanishes on odd elements $F$. This implies that 
$\gamma$  can be unitarily implemented in the GNS-representation 
of any quasifree state. In other words, there is a unitary $U$ on $\cF_\omega$ such that
\begin{equation}
U |\Omega_\omega\rangle = |\Omega_\omega \rangle, \quad \pi_\omega(\gamma(F)) = 
U \pi_\omega(F) U^* \quad \forall F \in \fF. 
\end{equation} 
One can then define the ``twisted (local) algebras'', $\pi_\omega(\fF(\cO))^t$, by 
\begin{equation}
\pi_\omega(\fF(\cO))^t = \{ \pi_\omega(F_+) + i\pi_\omega(F_-)U \mid F \in \fF(\cO)\}.
\end{equation}
The twisted algebra $\pi_\omega(\fF(\cO))^t$ is spatially isomorphic to $\pi_\omega(\fF(\cO))$ and 
the isomorphism is implemented by the unitary
$ V = \frac{1}{\sqrt{2}}(1 + i U)$

In terms of the twisted algebras, one has the following modified version of 
locality, called ``twisted locality'': 
\begin{equation}
\pi_\omega(\fF(\cO')) \subset \pi_\omega(\fF(\cO))^{\prime t}. 
\end{equation}
Here, $\pi_\omega(\fF(\cO))'$ denotes the commutant, that is, the set of all operators 
on $\cF_\omega$ that commute with all $\pi_\omega(F)$, $F \in \fF(\cO)$. The set 
$\cO'$ is the causal complement of the set $\cO$, defined as the set of all 
spacetime points $x$ such that $J(x) \cap \cO = \emptyset$. 

\medskip

The above construction of 
Fock states is appropriate to obtain ``ground states'' 
for static spacetimes $(M, g_{ab})$, which is the situation that we are
considering in Sec.~\ref{sec3}. Recall that a spacetime 
is called static if it has  
an everywhere time-like Killing vectorfield $t^a$, 
$\pounds_t \g_{ab} = 2 \nabla_{(a} t_{b)} = 0$, 
which is orthogonal to a family of Cauchy surfaces $\Sigma$, 
or, equivalently, which satisfies $t_{[a} \nabla_b t_{c]} = 0$. We 
also assume here that the Killing field is timelike everywhere in 
the sense that $v^2 = t^a t_a \ge v_0^2 > 0$ for some $v_0$, and that the 
orbits of $t^a$ are complete. The construction of the ground state of the Dirac field for such a spacetime is  
precisely as follows: Consider a Cauchy surface $\Sigma$ 
orthogonal to $t^a$. Since the Dirac equation has a well-posed initial
value formulation, we may associate with any given $k \in 
C^\infty_0(DM \restriction \Sigma)$ a uniquely determined solution $u$ of 
the Dirac equation such that $u \restriction \Sigma = k$. The flow 
$\{F_t\}_{t \in \mr}$ of $t^a$ induces a flow on the spinor bundle $DM$
(denoted by the same symbol), which commutes with the charge conjugation\footnote{
That can be seen e.g. as follows: Since the orbits of $t^a$ are complete and 
since $(M, g_{ab})$ is assumed globally hyperbolic, every point $x \in M$ 
can be written uniquely as $x = F_t(y), y \in \Sigma, t \in \mr$. Since
$F^*_t g_{ab} = g_{ab}$, we may 
assume that the set of spin-frames over any given point $x = F_t(y)$ is 
{\em defined} to be the set of spin-frames over the point $y$, and that $D_x M$ 
is consequently defined to be $D_y M$. This provides the desired (trivial) flow on 
the spin-bundle induced by $F_t$. The charge conjugation acting in the fiber
$D_x M$ is defined to be the charge conjugation in $D_y M$ and therefore (trivially)
commutes with the flow $F_t$.
} 
and the Dirac operator~\cite{dim1}. Therefore,  
$F^*_t u$ is again a solution to the Dirac equation with initial data
$k_t = F^*_t u \restriction \Sigma$. It is not difficult to see 
that $k_t = {\bf u}(t)k$ for some 1--parameter group of unitary 
maps $\{{\bf u}(t)\}_{t \in \mr}$ on $\cK$ satisfying 
$[{\bf u}(t), C] = 0$ for all $t \in \mr$. 
Hence, there is a 1-parameter family of Boguliubov 
automorphisms $\{\alpha_t\}_{t \in \mr}$ on $\fF$, given by 
$\alpha_t(\psi(k)) = \psi({\bf u}(t)k)$ for all $k \in \cK$. It is 
known \cite[Prop. 4.1]{strohm} that the 1--parameter group $\{{\bf u}(t)\}_{t \in \mr}$ is 
strongly continuous on $\cK$. Thus, by Stone's theorem, we can write 
\begin{equation}
{\bf u}(t) = e^{it{\bf h}}
\end{equation}
for some self-adjoint generator $\bf h$ on $\cK$. It can be seen the action of $\bf h$ on a
$k \in C^\infty_0(DM \restriction \Sigma)$ is given by the differential operator
\begin{equation}
\label{hdef}
{\bf h}k = (-i t^c \gamma_c q^{ab}\gamma_a \nabla_b + t^c \gamma_c m)k,
\end{equation}
where $q_{ab} = \g_{ab} - t_a t_b/v^2$ is the induced metric 
on $\Sigma$. Let now $P$ be the projector on the positive 
energy subspace of $\bf h$, 
\begin{eqnarray}
\label{Pdef}
P = \int_0^\infty E_{\bf h}({\rm d} \lambda),     
\end{eqnarray}
with $E_{\bf h}$ the spectral measure of ${\bf h}$. 
Since $C{\bf h}C = -{\bf h}$, the projector $P$ satisfies
$CPC = \myid - P$, and therefore defines
a Fock state $\omega$ on $\fF$. That Fock state is, in 
fact, the ground state associated with the notion of 
time translations defined by the Killing vector field $t^a$.

The group $\{\alpha_t\}_{t \in \mr}$ is unitarily implemented 
in the GNS-representation associated with the state $\omega$ in 
the sense that 
\begin{equation}
\pi_\omega(\alpha_t(F)) = {U}(t) \pi_\omega(F) 
{U}(t)^* \quad \text{for all 
$F \in \fF, t \in \mr$,} 
\end{equation}
where ${U}(t) = e^{itH}$ and where  
$H$ is given by the second quantization of the self 
adjoint operator (the ``1--particle Hamiltonian'') 
${\bf h}_+ = P {\bf h} P$. $H$ is the 
generator of the symmetry $t^a$ on 
the GNS Hilbert space associated with $\omega$ and may 
therefore be identified with the Hamiltonian of the system. 
The vacuum vector $|\Omega_\omega\rangle$   
is invariant under ${U}(t)$, i.e. $H |\Omega_\omega \rangle = 0$, 
thereby justifying the terminology ``ground state'' for $\omega$.  
In Minkowski space, the above ground state construction 
gives the usual vacuum representation. 

\subsection{Hadamard states}

In the previous subsection, we have recalled the definition of the algebra
of observables for a quantized Dirac field on a curved spacetime, and 
we have introduced the notion of quasifree states on this algebra. We also 
recalled the construction of the ground state in a static spacetime, which  
is a preferred element in this class. 
The question then arises how to characterize a class of physically reasonable
states on non-static, general globally hyperbolic spacetimes. 
It has been suggested for a long time 
that this class should include the quasifree 
states that have a two-point function of {\it Hadamard form}. 
As was recently shown by Hollands and Ruan  \cite{hr}, this is 
a necessary and sufficient condition in order to be able to 
extend the action of a quasifree state to more singular quantities such 
as Wick powers of the free field (for example stress-energy 
tensor)---which are not 
already contained in $\fF$---as well as to other quantities arising
in a perturbatively defined interacting quantum field theory.\footnote{
More precisely in \cite{hr}, the following general result
was proven for a Hermitian scalar field: If the algebra 
$\fF$ of free quantum fields is 
enlarged to a suitable algebra $\cW$ containing the observables of interest
in the perturbatively defined interacting theory, then the states
on $\fF$ that can be extended to $\cW$ are precisely the Hadamard
states with smooth truncated $n$-point functions for $n \neq 0$. 
This class includes in particular the quasifree Hadamard states, 
since these have vanishing truncated $n$-point functions for $n \neq 0$. 
The generalization of 
this result to Dirac fields is straightforward.} 
We will see in Sec.~\ref{sec4} that the ground state 
on an ultrastatic spacetime (that is, a static spacetime 
for which the timelike Killing field is in addition normalized, $t^a t_a = -1$)
is of Hadamard form. 

For the convenience of the reader, we now recall the definition 
of Hadamard states for the Majorana-Dirac field. Roughly speaking, 
these are states whose two-point function has the short distance
singularity structure of a Hadamard fundamental solution
of the Dirac equation with no spacelike singularities. 
A mathematically precise definition of Hadamard states was 
first given by Kay and Wald \cite{kw} for the case of 
a scalar field and that definition has subsequently been generalized
to Dirac fields by \cite{ver4}. We here prefer to work with 
an alternative equivalent characterization of Hadamard 
states in terms 
of the so-called ``wave front set'' of their associated two-point function
(for a definition of this concept see \cite{ho}).  
This characterization was 
first found by Radzikowski \cite{r} in the case of scalar fields, and 
was later also established for Dirac fields, see \cite{h, kk, sv1}. 
\begin{defn}
\label{hadadef}
A state $\omega$ on $\fF$ is said to be of Hadamard form if the 
wave front set, ${\rm WF}({\cal S}_\omega)$, of its two-point function is 
contained in the set
\begin{equation}
C^+ = \{(x_1, \xi_1; x_2, -\xi_2) \in T^* M^2 \setminus \{0\} \mid
(x_1, \xi_1) \sim (x_2, \xi_2), \xi_1 \triangleright 0\}. 
\end{equation}
Here, the notation $(x_1, \xi_1) \sim (x_2, \xi_2)$ means that (a) the points
$x_1$ and $x_2$ can be joined by a null-geodesic $c: [0,1] \to M$, 
(b) the covectors $\xi_1$ and $\xi_2$ are cotangent to $c$ (meaning that 
$\dot c^a(0) = g^{ab}(x_1)\xi_{1b}$ and $\dot c^a(1) = 
g^{ab}(x_2)\xi_{2b}$) and parallel transports of each other. 
$\xi_1 \triangleright 0$ means that $\xi_1$ is future-pointing.
\end{defn}
\paragraph{Remarks.}
1) An important consequence of the above definition  
is that if $\omega_1$ and $\omega_2$ are two Hadamard states, 
then the distribution ${\cal K} = {\cal S}_{\omega_1} - 
{\cal S}_{\omega_2}$ is given by a smooth function. To see this, we 
notice that 
\begin{equation}
{\cal K}(Cu_1, Cu_2) = -\overline{{\cal K}(u_1, u_2)}, 
\end{equation}
by the Hermiticity and the commutation relations of the 
Dirac field. Since $C$ is anti-linear, this equation implies, together 
with the definition of the wave front set, that 
if $(x_1, \xi_1; x_2, \xi_2) \in {\rm WF}({\cal K})$, then 
also $(x_1, -\xi_1; x_2, -\xi_2) \in {\rm WF}({\cal K})$; in other words
\begin{equation}
{\rm WF}({\cal K}) = -{\rm WF}({\cal K}).
\end{equation}
On the other hand, we know that 
\begin{equation}
{\rm WF}({\cal K}) = {\rm WF}({\cal S}_{\omega_1} - 
{\cal S}_{\omega_2}) \subset {\rm WF}({\cal S}_{\omega_1})
\cup {\rm WF}({\cal S}_{\omega_2}) \subset C^+, 
\end{equation}
because $\omega_1$ and $\omega_2$ are Hadamard. Since the 
only subset of $S \subset C^+$ with $S = -S$ is the 
empty set, this implies that in fact ${\rm WF}({\cal K}) = 
\emptyset$, which is equivalent to the statement that $\cal K$
is smooth.  

\smallskip
\noindent
2) If a quasifree state is of Hadamard form in a globally hyperbolic neighbourhood of 
a Cauchy surface, then, as originally shown by Fulling, Narcowich and Wald \cite{fnw}, 
it must in fact be of Hadamard form throughout the whole 
spacetime. For a more recent version of this argument within the framework 
of microlocal analysis, see for example \cite{h, kk, sv1}. 

\section{Local quasiequivalence and factoriality}
\label{sec3.1}

We will now show that the GNS representations 
$\pi_{\omega_1}$ and $\pi_{\omega_2}$ of any two quasifree Hadamard 
states $\omega_1$ and $\omega_2$ on $\fF$ are quasiequivalent when restricted to $\fF(\cO)$
for any open, relatively compact set $\cO \subset M$, meaning 
that the density matrices in the GNS Hilbert spaces $\cF_{\omega_1}$ and $\cF_{\omega_2}$ 
define the same sets of states of $\fF(\cO)$.

In order to prove this result, we will proceed as follows: 
We first note that $\pi_{\omega_1} \restriction \fF(\cO)$ 
is quasiequivalent to $\pi_{\omega_2} \restriction \fF(\cO)$, if
$\pi_{\omega_1} \restriction \fF(\tilde \cO)$ 
is quasiequivalent to $\pi_{\omega_2} \restriction \fF(\tilde \cO)$
for some $\tilde \cO \supset \cO$. Now
every open, relatively compact set $\cO \subset M$ can be embedded
into a set of the form ${\rm int} (D(\cC))$, where $\cC$ is a
relatively compact subset of some Cauchy surface $\Sigma$ with smooth boundary in $\Sigma$~\footnote{
Proof: Let $K$ be a compact set containing $\cO$, and let $\cC = (J^+(K) \cap \Sigma) \cup 
(J^-(K) \cap \Sigma)$. Since $\Sigma$ is a Cauchy surface, it follows that 
$D(\cC) \supset K$ by the definition of the domain of dependence. 
We will show that $\cC$ is compact by showing that $J^\pm(K) \cap \Sigma$ is compact. The collection of 
sets $\{J^-(x)\}_{x \in M}$ is a cover of $K$ and hence has a finite subcover $\{J^-(x_1), 
\dots, J^-(x_r) \}$. By Thm.~8.3.12 of~\cite{w}, $J^-(x) \cap \Sigma$ is compact for any point $x$,
and therefore so is also $\cup_{i=1}^r (J^-(x_i) \cap \Sigma)$. Hence, $J^-(K) \cap \Sigma$ is 
contained in a compact set. But $J^-(K) \cap \Sigma$ is also closed by Thm.~8.3.11 of~\cite{w}, hence compact. 
The same argument can be repeated for ``$+$'', showing that $\cC$ is a compact subset of $\Sigma$. 
By enlarging $\cC$ if necessary, we can achieve that it has a smooth boundary in $\Sigma$.}. We 
may therefore restrict attention to regions $\cO_\cC$ of the form ${\rm int} (D(\cC))$. 
For such regions, we first show that the GNS representation
$\pi_{\omega \restriction \fF(\cO_\cC)}$ of the partial state $\omega \restriction
\fF(\cO_\cC)$ is quasiequivalent to the representation $\pi_\omega \restriction \fF(\cO_\cC)$. 
Then we will check that the representations relative to any partial (quasifree, Hadamard) 
state $\omega \restriction \fF(\cO_\cC)$ are locally quasiequivalent. 

\begin{lem}
\label{lemB}
Let $A \supset B$ be $C^*$-algebras, $\omega$ a state on $A$, and $(\pi, \cF, |\Omega\rangle)$
be the GNS representation of $A$. Denote by $e'$ the orthogonal projection on 
$\overline{\pi(B) | \Omega \rangle}$. Then $\pi_{\omega \restriction B}$ is 
quasiequivalent to $\pi_\omega \restriction B$ if and only if the central support of 
$e'$ is $\myid$. 
\end{lem}
\begin{proof}
The projection $e'$ belongs to $\pi(B)'$ and $B \owns b \to \pi(b) \restriction e'\cF$ is
equivalent to the GNS representation of $\omega \restriction B$. If the central support 
of $e'$ in $\pi(B)'' \cap \pi(B)'$ is equal to $\myid$, then the homomorphism (induction)
$\pi(B)'' \owns T \to T \restriction e' \cF$ is an isomorphism that gives the 
quasiequivalence between $B \owns b \to \pi(b)$ and $B \owns b \to \pi(b) \restriction e'\cF$.
Conversely, quasiequivalent representations have equivalent amplifications. 
\end{proof}

In order to be able to apply the above lemma in the case at hand we next prove 
a proposition that is also of some interest in its own right. 

\begin{prop}
\label{factoriality}
Let $\cC$ be a relatively compact subset of a Cauchy surface $\Sigma$ with smooth boundary 
in $\Sigma$, and let $\cO_\cC = {\rm int}(D(\cC))$, where $D(\cC)$ is the domain of dependence 
of $\cC$. Then, for any quasifree state $\omega$,
\begin{equation}
\label{td}
\pi_\omega(\fF(\cO_\cC))^{\prime t} \cap \pi_\omega(\fF(\cO_\cC))'' = \mc \cdot \myid.
\end{equation}  
\end{prop}
\paragraph{\bf Remark:}
Taking into account the even and odd parts $\fF_\pm$ separately, the proposition implies that 
the  even part of the center of $\pi_\omega(\fF(\cO_\cC))$ is trivial. 

\begin{proof}
Let $P$ be the operator on $\cK = L^2(DM \restriction \Sigma)$ corresponding to the quasifree 
state $\omega$ via eq.~\eqref{fst}, and let $\cH \subset \cK$ be the 1-particle Hilbert space, 
as defined in eq.~\eqref{Hdef}. In addition, let us define the following closed, 
real linear subspaces of $\cH$: 
\begin{eqnarray*}
\cM(\cC) &=& {\rm clo} \{ P^{1/2} k \mid k \in C^\infty_0(DM \restriction \cC), \quad Ck = k\},\\ 
\cM      &=& {\rm clo} \{ P^{1/2} k \mid k \in C^\infty_0(DM \restriction \Sigma), \quad Ck = k\}. 
\end{eqnarray*}
We note that $\cH = \cM + i \cM$ and it is not difficult to see that $\cH = \cM$ if and only if 
$P$ is a projector. By eq.~\eqref{repdef}, we can write
\begin{equation*}
\pi_\omega(\fF(\cO_\cC))'' = \{ a(m) + a(m)^* \mid m \in \cM(\cC)\}''.
\end{equation*}  
By an argument due to Foit \cite{foit} and Roberts, we know that \eqref{td} holds if and only if
$\cM(\cC) \cap i\cM(\cC)' = \{0\}$, where ``prime'' denotes the symplectic complement of 
a set in $\cH$, defined as 
\begin{equation}
\cM(\cC)' = \{ m' \in \cH \mid {\rm Im} (m | m') = 0 \quad \forall m \in \cM(\cC) \}.)
\end{equation}
Let now $m \in \cM(\cC) \cap i\cM(\cC)'$. We must show that $m = 0$. Now, 
since $m \in i\cM(\cC)'$, we have that $im \in \cM(\cC)'$, which means 
that ${\rm Re}(m | m') = 0$ for all $m' \in \cM(\cC)$.
Moreover, since $m \in \cM(\cC)$, it is not difficult to see that ${\rm Re}(m | m') = 0$
for all $m' \in \cM(\Sigma \setminus \cC)$. Altogether, this means that 
\begin{equation}
{\rm Re}(m | m') = 0 \quad \forall m' \in \cM(\cC) + \cM(\Sigma \setminus \cC).
\end{equation} 
Consequently, we know that 
\begin{equation}
{\rm Re}(m | P^{1/2} k') = 0 \quad \forall k' \in C^\infty_0(DM \restriction \cC) + C^\infty_0
(DM \restriction \Sigma \setminus \cC), \,\, Ck' = k'.
\end{equation} 
Because the boundary of $\cC$ is smooth, it follows easily that the space 
$C^\infty_0(DM \restriction \cC) + C^\infty_0(DM \restriction \Sigma \setminus \cC)$ is 
dense in $C^\infty_0(DM \restriction \Sigma)$ in 
the norm induced by the scalar product \eqref{ip}. This implies that 
\begin{equation}
\label{mm}
{\rm Re}(m | m') = 0 \quad \forall 
m' \in \cM, 
\end{equation} 
or equivalently that $m \in \cM \cap i\cM'$. Let us now introduce
an auxiliary quasifree state $\tilde \omega$ whose corresponding operator $\tilde P$ is a projection, 
so that $\tilde \cH = \tilde \cM$ holds for the corresponding 1-particle space. 
Let us write $m = \lim_{j \to \infty} P^{1/2} k_j$, where $k_j \in 
C^\infty_0(DM \restriction \Sigma)$ with $Ck_j = k_j$. Then, since 
\begin{equation}
\label{normeq}
\|P^{1/2} k_i - P^{1/2} k_j\| = \|k_i - k_j\|
= \|\tilde P^{1/2} k_i - \tilde P^{1/2} k_j\|,  
\end{equation}
it follows that the limit $\tilde m = \lim_{j \to \infty} \tilde P^{1/2} k_j$ exists in 
$\tilde \cM$. If $m' = P^{1/2} k'$, with $k'$ 
an arbitrary element of $C^\infty_0(DM \restriction \Sigma)$ with $Ck' = k'$, 
then \eqref{mm} implies that 
\begin{eqnarray*}
0 &=& {\rm Re} (m|m') \\
&=& \lim_{j \to \infty} {\rm Re} (P^{1/2} k_j |P^{1/2} k')\\
&=& \lim_{j \to \infty} {\rm Re} (\tilde P^{1/2} k_j | \tilde P^{1/2} k')\\
&=& {\rm Re} (\tilde m|\tilde m'), 
\end{eqnarray*}
where $\tilde m' = \tilde P^{1/2} k'$. It follows from this that ${\rm Re} (\tilde m|\tilde m')$
for all $\tilde m' \in \tilde \cM = \tilde \cH$, and therefore that $\tilde m = 0$. But, by 
\eqref{normeq}, this also implies that $m = 0$, which proves the proposition. 
\end{proof}

We now will now combine the above two results to show
\begin{prop}
Let $\omega$ be a quasifree state on $\fF$. Then the representations $\pi_{\omega \restriction
\fF(\cO_\cC)}$ and $\pi_\omega \restriction \fF(\cO_\cC)$ are quasiequivalent. 
\end{prop}
\begin{proof}
Let $U$ be the unitary operator on $\cF_\omega$ implementing the 
automorphism $\gamma$, 
$\pi_\omega(\gamma(F)) = U\pi_\omega(F)U^*$. By the remark following 
Prop.~\ref{factoriality}, we know that the ${\rm Ad}(U)$-invariant 
(that is, even) part of 
of $\pi_\omega(\fF(\cO_\cC))'' \cap \pi_\omega(\fF(\cO_\cC))'$ is trivial. The central
support of $\overline{\pi_\omega(\fF(\cO_\cC))|\Omega_\omega\rangle}$ is 
invariant under ${\rm Ad}(U)$, hence it is $\myid$. The result now follows from 
Lem.~\ref{lemB}. 
\end{proof}
 
\begin{thm}
\label{lqe}
(Local quasiequivalence)
The GNS representations $\pi_{\omega_1}$ and $\pi_{\omega_2}$ 
of any two quasifree Hadamard states 
$\omega_1$ and $\omega_2$ are locally quasiequivalent when 
restricted to $\fF(\cO)$ for any open, relatively compact set $\cO \subset M$.
\end{thm}
\begin{proof}
As explained above, we only need to prove the quasiequivalence of 
the representations $\pi_{\omega_1 \restriction \fF(\cO_\cC)}$ 
and $\pi_{\omega_2 \restriction \fF(\cO_\cC)}$, 
where $\cO_\cC$ is a region of the form ${\rm int}(D(\cC))$.   
Let $P_1$ and $P_2$ be the self-adjoint operators on $\cK = L^2(DM \restriction \Sigma)$
associated to the the quasifree states $\omega_1$ and $\omega_2$ via 
eq.~\eqref{fst}. The partial states $\omega_1 \restriction \fF(\cO_\cC)$ and
$\omega_2 \restriction \fF(\cO_\cC)$ are then again quasifree states, corresponding to the 
operators $E_\cC P_1 E_\cC$ and $E_\cC P_2 E_\cC$, 
where $E_\cC$ is the projector onto the closed subspace 
$L^2(DM\restriction \cC)$ of $\cK$, given 
by $(E_\cC k)(x) = \chi_\cC(x) k(x)$ with $\chi_\cC$ the characteristic 
function of the set $\cC$. 

By a theorem of Powers and St\o rmer \cite[Thm. 5.1]{ps}, 
the partial states $\omega_1 \restriction \fF(\cO_\cC)$ and  
$\omega_2 \restriction \fF(\cO_\cC)$ 
are guaranteed to be quasiequivalent if 
\begin{equation}
\label{ecp}
\|(E_\cC P_1 E_\cC)^{1/2} - (E_\cC P_2 E_\cC)^{1/2}\|_2 < \infty,  
\end{equation}
where $\| \,\cdot \,\|_2$ is the Hilbert-Schmidt norm on 
$\cK$, see Sec.~\ref{sec2.2}. In order to estimate the 
left side of \eqref{ecp}, we recall the 
the Powers-St\o rmer inequality, which says that $\|A^{1/2} - B^{1/2}\|_2^2 
\le \|A - B\|_1$ for any two positive bounded operators $A$ and $B$ on 
a Hilbert space, where $\|\,\cdot\,\|_1$ denotes the trace-norm, 
see Sec.~\ref{sec2.2}.
Taking $E_\cC P_1 E_\cC$ for $A$ and 
$E_\cC P_2 E_\cC$ for $B$ in that inequality, we find that \eqref{ecp} follows
if we can show that 
\begin{equation}
\label{trcl}
\|E_\cC(P_1 - P_2)E_\cC\|_1 < \infty.
\end{equation}
We now define a bidistribution $K$ on $\cD(DM \restriction \Sigma) 
\times \cD(DM \restriction \Sigma)$ associated with the operator 
$E_\cC(P_1 - P_2)E_\cC$ by putting $K(k_1, k_2) 
= (Ck_1 | E_{\cC}(P_1 - P_2)E_{\cC}k_2)$. This bidistribution
can be expressed in terms of the two-point 
functions ${\cal S}_{\omega_1}$ and ${\cal S}_{\omega_2}$ as follows: 
By the Hadamard property
combined with a microlocal argument (see e.g. \cite{h}), 
the latter possess a well-defined pull-back to the 
submanifold $\Sigma \times \Sigma$, denoted by 
$\varphi_\Sigma^* {\cal S}_{\omega_1}$ respectively 
$\varphi_\Sigma^* {\cal S}_{\omega_2}$, where $\varphi_\Sigma: 
\Sigma \times \Sigma \to M \times M$ is the embedding map. 
Using eq.~\eqref{fst}, we compute that 
\begin{eqnarray}
\varphi_\Sigma^* {\cal S}_{\omega_j}(\gamma_a n^a k_1, \gamma_b n^b k_2)
&=& \left(C\rho_\Sigma^{} {\cal S} \rho_\Sigma' (\gamma_a n^a k_1)| 
P_j \rho_\Sigma^{} {\cal S} \rho_\Sigma' (\gamma_b n^b k_2) \right)\nonumber
\\
&=& ( Ck_1 | P_j k_2 ) \quad (j=1, 2),
\end{eqnarray}  
where $n^a$ is the normal of $\Sigma$ and where 
we have used the identity 
$\rho_\Sigma^{} {\cal S} \rho_\Sigma' = i\gamma_a n^a \myid_\Sigma^{}$ (see \cite{dim1})
in the last line, with $\rho'_\Sigma$ the dual of the restriction map $\rho_\Sigma^{}$. 
It is not difficult to see from this that 
\begin{equation}
\label{Kexp}
K(k_1, k_2) =
\varphi_\Sigma^*({\cal S}_{\omega_1} - 
{\cal S}_{\omega_2})(\chi_\cC \gamma_a n^a k_1, 
\chi_\cC \gamma_b n^b k_2) 
\end{equation} 
Since $\omega_1$ and $\omega_2$ are by assumption
Hadamard states, we know that ${\cal S}_{\omega_1} - {\cal S}_{\omega_2}$ is 
smooth on $M \times M$ (see the first remark following Def.~\ref{hadadef}). 
Equation~\eqref{Kexp} therefore implies that $K$ can be identified with a smooth section in 
$D^* M \restriction \Sigma \times D^*M \restriction \Sigma$, with compact support in 
$\cC \times \cC$. Let $\{ \eta_i \}$ be a smooth partition of unity covering $\cC$ 
with the property that the support of each $\eta_i$ is contained in a 
coordinate patch. Then we can write $K = \sum (\eta_i \otimes \eta_j) K = \sum K_{ij}$, 
and each $K_{ij}$ can be identified a matrix of smooth, compactly
supported functions in $\cD (\mr^{s} \times \mr^{s})$ via a local trivialization 
of $D^*M \restriction \Sigma \times D^*M \restriction \Sigma$ 
over the patch corresponding to the pair $\eta_i, \eta_j$. As it is well known, 
a matrix of operators is in the trace class if each 
matrix entry is in the trace class. Thus, in order to show that $K$ is 
in the trace class, it suffices to show that the matrix elements
of each $K_{ij}$ is in the trace class. In fact any smooth, compactly supported 
integral kernel $L$ in $\cD (\mr^{s} \times \mr^{s})$ is in the trace class, as
one can see e.g. as follows: We can write 
\ben
\label{lwrite}
L(x,y) = (1 + |x|)^{-p} (1 + |x|)^p L(x,y),  
\een
where $|x| = (\sum x_i^2)^{1/2}$. The multiplication operator by $(1 + |x|)^{-p}$ is 
in the Hilbert-Schmidt class 
for $p$ sufficiently large since $\|(1 + |x|)^{-p} \|_2^2 = \int(1 + |x|)^{-2p} \, d^{s} x < \infty$ then, 
and the integral kernel $(1 + |x|)^p L(x,y)$ is 
in $L^2(\mr^{s} \times \mr^{s})$ for any $p$, and therefore also in the Hilbert-Schmidt
class. Hence, $L$ is the product of two Hilbert-Schmidt operators, and hence in the trace
class.
\end{proof}

We conclude this section with the following theorem. 

\begin{thm}
\label{lf}
(Local factoriality)
Let $\omega$ be a quasifree state Hadamard state for the free Dirac field 
on a globally hyperbolic spacetime. Then the local v.~Neumann 
algebras $\pi_\omega(\fF(\cO_\cC))''$ are type $III_1$ factors. Furthermore, 
the local quasiequivalence established in Thm.~\ref{lqe} is in fact local 
unitary equivalence.
\end{thm}
\begin{proof}
The partial state $\omega \restriction \fF(\cO_\cC)$ is a quasifree state
corresponding to $E_\cC P E_\cC$ in the notation of the proof of Thm.~\ref{lqe}.
Hence, the algebra $\pi_{\omega \restriction \fF(\cO_\cC)}(\fF(\cO_\cC))$ is 
a factor by the results of Powers and St\o rmer \cite{ps}. By local quasiequivalence, 
also $\pi_\omega(\fF(\cO_\cC))$ is a factor. The type of this factor can 
be obtained by again invoking local quasiequivalence. In fact, the local 
v.~Neumann algebras are type $III_1$ factors in a pure, quasifree Hadamard
state by \cite[Cor. 5.12]{ver4}, therefore they are $III_1$ factors for all 
quasifree Hadamard states.  
\end{proof}

\paragraph{Remark.} By a similar argument, one can also show that the local 
algebras of observables, $\pi_\omega(\cA(\cO_\cC))''$ are factors of type $III_1$.

\section{Nuclearity}
\label{sec3}

In this section, we show that the theory of a free Dirac field on a static 
spacetime satisfies the nuclearity condition stated in the introduction. 
Before giving the precise formulation of our result, 
let us begin by fixing our conventions for this section, as well 
as some notations: We assume that the spacetime $(M, g_{ab})$ under consideration 
is static. Let $\Sigma_t$ be the Cauchy surfaces orthogonal to the timelike
Killing vector field. We pick an arbitrary but fixed 
$\Sigma = \Sigma_t$, and fix a subset $\cC \subset \Sigma$ 
with compact closure. For that subset, we define $\cO_\cC = {\rm int}(D(\cC)) \subset M$, 
where $D(\cC)$ is the domain of dependence of $\cC$. Throughout, we will work in 
the representation of $\fF$ given by the ground state 
$\omega$ introduced in Sec.~\ref{sec2.1}. In this representation, 
the Fermi fields take the form
\ben
\psi(k) = a(Pk)^* + a(PCk), \quad k \in \cK = L^2(\Sigma; DM), 
\een
where we are writing $\psi(k)$ instead of $\pi_\omega(\psi(k))$ 
to lighten the notation in this section. Here, 
$P$ is the projector on the positive spectral subspace of ${\bf h}$
(see eq.~\eqref{Pdef}), and $a(p)^*$ and $a(p)$ are the creation resp. annihilation operators 
defined in eq.~\eqref{adef}, with $p \in \cH = P\cK$. The local algebra $\fF(\cO_\cC)$
is then generated by those $\psi(k)$ for which $k$ is supported in $\cC$, that is
$k \in L^2(\cC; DM)$. It is viewed as 
a subalgebra of the algebra of bounded operators on $\cF_\omega$ via $\pi_\omega$. 
For further notational simplicity, we will also drop the subscript 
``$\omega$'' on $\cF_\omega$ and the vacuum vector $|\Omega_\omega\rangle$
in this section, the understanding being that we always refer to the 
quantities associated with the ground state. For the relevant definitions 
concerning the nuclearity of a map and related functional analytic 
concepts we refer the reader to Sec.~\ref{sec2.2}.

\medskip

Let $\Theta_\beta$ be the map defined by  
\begin{eqnarray}
\label{thetadef}
\Theta_\beta: \fF(\cO_\cC) \to \cF, \quad \Theta_\beta(A) 
= e^{-\beta H} A|\Omega\rangle, 
\end{eqnarray}  
for $\beta > 0$, and let $$\nu_{\cC, \beta} = \|\Theta_\beta\|_1$$ be its nuclearity index.  
The aim of the present section is to show that $\Theta_\beta$ is $p$-nuclear
(Thm.~\ref{main}) and to estimate the nuclearity index 
$\nu_{\cC, \beta}$ (Prop.~\ref{prop}). 

\begin{thm}
\label{main}
(Nuclearity) The maps $\Theta_\beta$ are $p$-nuclear 
for all $\beta > 0$ and all $p > 0$, i.e., $\|\Theta_\beta\|_p < \infty$. 
\end{thm}
\noindent
{\it Proof:}
We first show that $p$--nuclearity of the maps $\Theta_\beta$ follows if,
for each $\beta > 0$, there is a positive operator $T$ on $\cH = P \cK$, with the following two properties:
\begin{enumerate}
\item[(a)] $T \in \cI_p(\cH)$ for all $p>0$ and all $\beta > 0$, 
where $\cI_p(\cH)$ denotes the Schatten-space with index $p$, consisting of 
all bounded operators $B$ such that $\| B \|_p = ({\rm tr} |B|^p)^{1/p} < \infty$
(see also Sec.~\ref{sec2.2}).  
\item[(b)] 
If $\{p_j\}_{j \in \mn}$ is an orthonormal basis
of $\cH$ consisting of eigenvectors of $T$ and 
$\{t_j\}_{j \in \mn}$ the corresponding eigenvalues, 
\begin{equation*}
Tp_j = t_j p_j, 
\end{equation*}
then there holds the inequality
\begin{eqnarray}
\label{*}
\left| \langle \Phi_{i_1 \dots i_k}| \Theta_\beta(A)\rangle \right|
\le t_{i_1} \dots t_{i_k} \|A\| \quad \text{for all $A \in \fF(\cO_\cC)$,}
\end{eqnarray}
where $|\Phi_{i_1 \dots i_k}\rangle = 
a(p_{i_1})^* \dots a(p_{i_k})^* |\Omega\rangle$ with $i_1 < \dots < i_k$.   
\end{enumerate}
To see that the existence of such an operator $T$ indeed implies the 
statement of the theorem, we expand $\Theta_\beta$ as  
\begin{eqnarray*}
\Theta_\beta(A) = \sum_{k \ge 0} \sum_{i_1 < \dots < i_k} 
|\Phi_{i_1 \dots i_k} \rangle \varphi_{i_1 \dots i_k}(A)
\end{eqnarray*}
where $\varphi_{i_1 \dots i_k}(A) = \langle \Phi_{i_1 \dots i_k}| \Theta_{\beta}(A) \rangle$.  
Since $\| | \Phi_{i_1 \dots i_k} \rangle \| = 1$ and since 
$\|\varphi_{i_1 \dots i_k}\| \le t_{i_1} \dots t_{i_k}$ by 
eq.~\eqref{*} in assumption (b), we get the estimate
\begin{eqnarray*}
\|\Theta_\beta\|_p &\le& \left(\sum_{k \ge 0} \sum_{i_1 < \dots < i_k}
\|\varphi_{i_1 \dots i_k}\|^p  \right)^{\frac{1}{p}} \\
&\le&\left(\sum_{k \ge 0} \sum_{i_1 < \dots < i_k}
t_{i_1}^p \dots t_{i_k}^p  \right)^{\frac{1}{p}} \\
&=& \left(\sum_{k \ge 0} \tr\left(\wedge^k(T^p)\right)  
\right)^{\frac{1}{p}} = (\det (\myid + T^p))^{\frac{1}{p}}, 
\end{eqnarray*}
where in the last line we have used a well-known identity for 
trace--class operators, see e.g. \cite[Thm.~XIII.106]{reed&simon}. 
Using the inequality~\cite[Lem.~4, Sec.~XIII.17]{reed&simon}, 
\begin{eqnarray*}
|\det(\myid + T^p)| \le e^{\|T^p\|_1}, 
\end{eqnarray*}
we conclude that
\begin{eqnarray}
\label{thetab}
\|\Theta_\beta\|_p \le e^{\frac{1}{p}\| T \|_p^p} 
\end{eqnarray}
which is finite, because $T \in \cI_p(\cH)$ by assumption (a). 
This then shows that the maps $\Theta_\beta$ 
are $p$-nuclear for any $p > 0$
and  $\beta > 0$, with $\|\Theta_\beta\|_p \le
(\det(\myid + T^p))^{1/p}$.

In order to complete the proof, we must show 
the existence of an operator $T$ with the above properties (a) and (b).  
Let $E_{\cC}$ be the projector onto the closed subspace 
$L^2(DM \restriction \cC)$ of $\cK$ (i.e., multiplication by the characteristic 
function of the set $\cC$), define  
\begin{equation}\label{Sdef}
S = 2 E_{\cC} e^{-\beta {\bf h}} P,  
\end{equation}
and let $S = VT$ be the polar decomposition of $S$ (so that 
$T = (S^* S)^{1/2}$). In what follows, we will establish that the so 
defined operator $T$ has the above stated properties (a) and (b). 
Let us begin by stating some relevant properties of $S$ and $T$.
\begin{lem}
\label{lem1}
The operators $T$ and $S$ have the following properties: 
\begin{enumerate}
\item[(i)] $T = PT = TP$ and $\|T\| \le \sqrt{2}e^{-\beta m_0}$.
\item[(ii)] $T \in \cI_p(\cH)$ for
all $p > 0$ for all $\beta > 0$.
\item[(iii)] There holds the estimate 
$$\|S\|_1 \le  \left(\frac{\beta_0}{\beta}\right)^s 
e^{-\beta m_0/2},$$ 
where $s$ is the number of spatial dimensions, $m_0 = v_0 m$ and
$\beta_0 > 0$ is a constant independent of $\beta$. 
\item[(iv)]
The range of $S^*$ is dense in $\cH$. 
\end{enumerate}
\end{lem}
\begin{proof}
The first part of (i) follows immediately from the definition. In 
order to show the norm estimate in (i), we notice that \eqref{hdef} 
implies that  
\begin{eqnarray}
\label{h2def}
{\bf h}^2 \ge m_0^2 \myid, 
\end{eqnarray}
where $m_0 = v_0 m$. We therefore get the inequality 
$e^{-\beta {\bf h}}P \le e^{-m_0\beta}$ and hence $\|T\| = \|T^2\|^{1/2} =
\sqrt{2} \|e^{-\beta {\bf h}} P E_\cC P e^{-\beta {\bf h}}  \|^{1/2} 
\le \sqrt{2} e^{-\beta m_0}$.  
Part (ii) of the lemma can be demonstrated by adapting a method developed 
in~\cite{dantoni&buchholz}, we briefly sketch the argument.
Let $\chi \in C^\infty_0(\Sigma)$ be a function which is 
equal to one on $\cC$ and let $M_\chi$ be the corresponding multiplication
operator on $\cK = L^2(DM \restriction \Sigma)$, defined by $M_\chi k = \chi k$. Then one can show that the operators 
\begin{eqnarray*}
{\bf k}_n = (\myid + \beta^2 {\bf h}^2)^{(n-1)s/2}M_\chi
(\myid + \beta^2 {\bf h}^2)^{-ns/2}   
\end{eqnarray*}
are in the Hilbert-Schmidt class for all $n \ge 1$. A proof of this is given
in the Appendix of ref.~\cite{buchholz&wichmann} for the 
case of Minkowski space; the main steps for a proof of this statement 
in the case when $\Sigma$ is not flat are given in the Appendix of this paper. 
Now it trivially follows from the definition~\eqref{Sdef} of $S$ that one can 
write
\begin{eqnarray*}
S = 2 E_{\cC} \cdot
{\bf k}_1 \cdot {\bf k}_2 \cdot \dots \cdot {\bf k}_n
(\myid + \beta^2{\bf h}^2)^{ns/2} e^{-\beta|{\bf h}|}P
\end{eqnarray*}
Since the operator $(\myid + \beta^2{\bf h}^2)^{ns/2} e^{-\beta|{\bf h}|}$
is bounded for any $\beta > 0, n \ge 0$, this shows that
$S$ can be written as the product of an arbitrary number of Hilbert--Schmidt
operators. This implies that $S \in \cI_p(\cK)$ for all $p > 0$ 
and hence $T \in \cI_p(\cH)$ for all $p > 0$, thus proving part 
(ii) of the lemma. The proof of (iii) is 
outlined in the Appendix. Part (iv) is equivalent to the 
statement that the range of the map $PE_\cC$ is dense in $\cH = P\cK$, 
which in turn is equivalent to the statement that the Fock state on $\fF$ 
given by $P$ has the ``Reeh-Schlieder-property''. That this property holds
in the situation under consideration has been shown in \cite{strohm}. 
\end{proof}

The above lemma establishes that $T$ has the desired property (a). It remains to 
be shown it also satisfies (b), i.e., that the eigenvalues $t_i$ of $T$ are related to the map $\Theta_\beta$
via eq.~\eqref{*}. The remainder of this proof consists in establishing this connection. 
For this let us first consider an arbitrary $A \in \fF$, and arbitrary $q_1, \dots, q_n \in 
\cH = P \cK$. Since $q_i = P q_i$, we have that
\ben
\psi(q_i) = a(q_i)^* + a(PCq_i) = a(q_i)^*, 
\een
because $PCq_i = PCPq_i = P(1-P)Cq_i = 0$, using that $CP = (1 - P)C$ and that $P^2 = P$.
Hence, we know that 
\ben
\label{f1}
a(q_1)^* \dots a(q_n)^* | \Omega \rangle = \psi(q_1) \dots \psi(q_n) | \Omega \rangle, 
\een
as well as
\ben
\label{f2}
\psi(q_i)^* | \Omega \rangle = a(q_i) | \Omega \rangle = 0. 
\een
Using the first relation~\eqref{f1}, we can write
\ben
\langle a(q_1)^* \dots a(q_n)^* \Omega | A \Omega \rangle = \langle \Omega | \psi(q_n)^* \dots \psi(q_1)^* A | \Omega \rangle.
\een
Applying the second relation~\eqref{f2}, we have
\ben
\label{f3}
\langle a(q_1)^* \dots a(q_n)^* \Omega | A \Omega \rangle = \langle 
\Omega | [ \psi(q_n)^* \dots [ \psi(q_2)^*, [\psi(q_1)^*, A]_\gamma ]_\gamma
 \dots ]_\gamma  \Omega \rangle, 
\een
because the terms arising from the multiple (graded) commutators having one or more factor of $\psi(q_i)^*$ standing
to the right of $A$ make no contribution on account of eq.~\eqref{f2}. We now use this relation to estimate 
the quantity $|\langle \Phi_{i_1 \dots i_k} | \Theta_\beta(A) \rangle |$, where $A \in \fF(\cO_\cC)$, and 
$|\Phi_{i_1 \dots i_k} \rangle = a(p_{i_1})^* \dots a(p_{i_k})^* | \Omega \rangle$. We have 
\begin{multline}
\label{e1}
\langle \Phi_{i_1 \dots i_k} | \Theta_\beta(A) \rangle = 
\langle a(p_{i_1})^* \dots a(p_{i_k})^* \Omega | e^{-\beta H} A\Omega \rangle = \\
\langle a(P e^{-\beta {\bf h}} p_{i_1})^* \dots a(P e^{-\beta {\bf h}} p_{i_k})^*\Omega |  
A \Omega \rangle,  
\end{multline}
where it was used that $p_i = Pp_i$ by definition. Next, apply eq.~\eqref{f3} with $q_i = P e^{-\beta {\bf h}} p_i$. This gives
\begin{multline}
\label{e2}
\langle \Phi_{i_1 \dots i_k} | \Theta_\beta(A) \rangle = \\
\langle \Omega | [ \psi(P e^{-\beta {\bf h}} p_{i_k})^* \dots [ \psi(P e^{-\beta {\bf h}} p_{i_2})^*, [\psi(
P e^{-\beta {\bf h}} p_{i_1})^*, A]_\gamma ]_\gamma
 \dots ]_\gamma  \Omega \rangle.  
\end{multline}
So far we have not used our assumption that 
$A$ is localized in the region $\cO_\cC$, i.e., $A \in \fF(\cO_\cC)$. In order to exploit this fact we write 
\ben
P e^{-\beta {\bf h}} p_{i} = E_\cC P e^{-\beta {\bf h}} p_{i} + (1 - E_\cC) P e^{-\beta {\bf h}} p_{i}, 
\een
where $E_\cC$ denotes multiplication by the characteristic function of the subset $\cC \subset \Sigma$. Inserting
this decomposition, we have
\begin{eqnarray}
[\psi(P e^{-\beta {\bf h}} p_{i_1})^*, A]_\gamma &=& 
[\psi((1-E_\cC)P e^{-\beta {\bf h}} p_{i_1})^*, A]_\gamma + [\psi(E_\cC P e^{-\beta {\bf h}} p_{i_1})^*, A]_\gamma\nonumber\\
&=& [\psi(E_\cC P e^{-\beta {\bf h}} p_{i_1})^*, A]_\gamma, 
\end{eqnarray}
where we have used in the second step that, due to graded locality~\eqref{gradloc}, $A$ has vanishing graded commutator with 
$\psi((1-E_\cC)P e^{-\beta {\bf h}} p_{i_1})^*$ because the latter is localized in the causal 
complement of $\cO_\cC$ by virtue of the multiplication by $1-E_\cC$, the characteristic function of 
$\Sigma \setminus \cC$. Thus, on the right side of eq.~\eqref{e2}, we are allowed to replace 
the expression $P e^{-\beta {\bf h}} p_{i_1}$ in the innermost commutator by the localized 
expression $E_\cC P e^{-\beta {\bf h}} p_{i_1}$. We now repeat this procedure for the next-to-innermost 
commutator and so fourth. After $k$ steps, we arrive at the identity
\begin{multline}
\label{e3}
\langle \Phi_{i_1 \dots i_k} | \Theta_\beta(A) \rangle = \\
\langle \Omega | [ \psi(E_\cC P e^{-\beta {\bf h}} p_{i_k})^* \dots [ \psi(E_\cC P e^{-\beta {\bf h}} p_{i_2})^*, [\psi(
E_\cC P e^{-\beta {\bf h}} p_{i_1})^*, A]_\gamma ]_\gamma
 \dots ]_\gamma  \Omega \rangle.  
\end{multline}
We can now substitute the definition of $S$ (see eq.~\eqref{Sdef}) into the right side of eq.~\eqref{e3}, giving
\ben
\label{e4}
\langle \Phi_{i_1 \dots i_k} | \Theta_\beta(A) \rangle = 
2^{-k} \langle \Omega | [ \psi(Sp_{i_k})^* \dots [ \psi(S p_{i_2})^*, [\psi(S p_{i_1})^*, A]_\gamma ]_\gamma
 \dots ]_\gamma  \Omega \rangle.  
\een
Using next that $S = VT$ and $Tp_{i} = t_i p_i$, we arrive at
\ben
\label{e5}
\langle \Phi_{i_1 \dots i_k} | \Theta_\beta(A) \rangle = 
2^{-k} t_{i_1} \dots t_{i_k} \langle \Omega | [ \psi(Vp_{i_k})^* \dots [ \psi(V p_{i_2})^*, [\psi(V p_{i_1})^*, A]_\gamma ]_\gamma
 \dots ]_\gamma  \Omega \rangle.  
\een
The right side can be estimated using the Cauchy-Schwarz inequality, giving 
\ben
\label{e6}
|\langle \Phi_{i_1 \dots i_k} | \Theta_\beta(A) \rangle| \le 
2^k \cdot 2^{-k} \cdot t_{i_1} \dots t_{i_k} \| A \| \prod_j^k \| \psi(Vp_{i_j}) \|,  
\een
where it was used that there are a total number of $2^k$ terms arising from the $k$ repeated 
commutators, each of which is estimated in the same fashion by the Cauchy-Schwarz inequality. We finally use 
the inequalities $\| \psi(k) \| \le \|k\|, k \in \cK$, and $\| V p_i \| \le \|V\| \, \| p_i \| = 1$, since $V$ is an 
isometry. This immediately gives $\| \psi(Vp_i) \| \le 1$, and thereby the desired inequality~\eqref{*}.

\qed

The proof of the theorem implies the following proposition.
\begin{prop}
\label{prop}
The $p$-norm of $\Theta_{\beta}: \fF(\cO_\cC) \to \cF$ is bounded by 
\begin{eqnarray}
\label{index} 
\|\Theta_\beta\|_p \le [\det(\myid + (S^* S)^{p/2})]^{1/p}, 
\quad \text{where $S = 2 E_{\cC} 
e^{-\beta {\bf h}} P$.}
\end{eqnarray}
Here, $E_{\cC}$ is the 
projection onto the closed subspace $L^2(DM \restriction \cC)$ of $\cK$,
$\bf h$ is the 1-particle Hamiltonian and $P$ projects onto the positive
spectral subspace of $\bf h$.
Furthermore, we have the following estimate for $\nu_{\cC, \beta}$:
\begin{eqnarray}
\label{nu1}
\nu_{\cC, \beta} = \|\Theta_\beta\|_1
\le \exp \left[\left(\frac{\beta_0}{\beta}\right)^s e^{-\beta m_0/2} \right] 
\end{eqnarray}
where $\beta_0 > 0$ is a constant depending only on the spatial geometry 
within $\cC$, $m_0 = v_0 m$ and $s$ is the number of spatial dimensions. 
If the the spacetime metric is rescaled by 
$g_{ab} \to \lambda^{2} g_{ab}$, then the 
nuclearity index for the rescaled spacetime 
(with Killing field $\lambda^{-1} t^a$) is estimated by the 
right side of \eqref{nu1}, with $\beta_0$ replaced by $\lambda \beta_0$.
\end{prop}
\begin{proof}
The inequality \eqref{index} is just eq. \eqref{thetab}. The estimate for $\nu_{\cC, \beta}$
follows from this and  (iii) of Lem. \ref{lem1}. In order to prove the last statement of the proposition, 
we note that the nuclearity index in 
the rescaled spacetime with Killing field $\lambda^{-1} t^a$ 
is equal to the nuclearity index in the unscaled 
spacetime at inverse temperature $\lambda^{-1}\beta$ and  
mass $\lambda m$. 
Since $v_0$ does not change under these rescalings, 
this means that the nuclearity index for the
rescaled spacetime is given estimated by the same expression 
as for the unscaled spacetime, but with $\beta_0$ replaced
by $\lambda \beta_0$. 
\end{proof}

\section{The split-property}
\label{sec4}
We are now going to show that the nuclearity property for Dirac fields in static spacetimes 
derived in the previous section implies the split-property. 

\begin{thm}\label{split}(Split-property)
Let $(M, g_{ab})$ be an arbitrary globally hyperbolic spacetime and $\omega$ a 
quasifree Hadamard state on $\fF$. Let $\cC_1$ and $\cC_2$ be two open subsets
of a Cauchy surface $\Sigma$ with ${\rm clo} (\cC_1) \subset \cC_2$ and set 
$\cO_1 = {\rm int}(D(\cC_1))$ and $\cO_2 = {\rm int}(D(\cC_2))$. Then there exists a 
type I factor $\cN$ on $\cF_\omega$ interpolating between the v.~Neumann algebras 
$\pi_\omega(\fF(\cO_1))''$ and $\pi_\omega(\fF(\cO_2))''$ in the sense that 
\begin{equation}
\pi_\omega(\fF(\cO_1))'' \subset \cN \subset \pi_\omega(\fF(\cO_2))''.
\end{equation}
\end{thm}
\paragraph{Remark:}
An analogous result holds for the algebras of observables $\cA(\cO) = \fF_+(\cO)$. 

\begin{proof}
Our proof follows the chain arguments given by Verch \cite[Prop. 5]{v} in the 
context of a linear Klein-Gordon field, thereby using the results obtained in the previous 
section and some results obtained in \cite{h}. 

Let us first consider the special case when $(M, g_{ab})$ is ultrastatic and 
when $\omega$ is the ground state. In that case we can define a map $\Theta_\beta$
as in eq.~\eqref{thetadef}, and we know by Thm.~\ref{main} that this map is nuclear with a nuclearity 
index estimated by the bound given in Prop.~\ref{prop}. It can be shown (see \cite[Thm. 2.1]{bdf}) that 
such a bound entails the split-property, thus proving our theorem in the case when $\omega$ is
the ground state in an ultrastatic spacetime. 

In order to generalize this statement from an ultrastatic spacetime to an arbitrary globally hyperbolic
spacetime (with spin-structure), one now makes use of the following facts:

\begin{enumerate}
\item[(i)]
The ground state on an ultrastatic spacetime is of Hadamard form. 
\item[(ii)]
More generally, if 
$(M, g_{ab})$ is a spacetime which is ultrastatic in an open 
neighborhood of some Cauchy surface $\Sigma$, then the state obtained from
the ``ground state construction on $\Sigma$'' is Hadamard throughout $M$.     
\end{enumerate}

Statement (i) is a corollary of the results obtained in \cite{sv}.
If the Cauchy surface $\Sigma$ is compact, then one can obtain the following alternative, 
somewhat more direct proof of (i) using a construction of Junker~\cite{ju}
(and using ideas of~\cite{h} to make Junker's construction applicable to the Dirac case). 
We first write the projector $P$ on the positive spectral subspace of $\bf h$ as 
$P = \frac{1}{2}|{\bf h}|^{-1}({\bf h} + |{\bf h}|)$. 
The operators ${\bf h}$ and $|{\bf h}|$ can be defined for every Cauchy surface 
$\Sigma$ perpendicular to the timelike Killing vector $t^a$, and can thereby act on 
smooth spinor fields $u$ on $M$. This makes it possible 
rewrite the two-point function \eqref{fst} of the ground 
state on an ultrastatic spacetime as 
\begin{eqnarray}
\label{possib}
{\cal S}_\omega(u_1, u_2) &=& 
\Big(C \rho_\Sigma{\cal S}u_1 \Big| 
P \rho_\Sigma {\cal S}u_2\Big) \nonumber \\
&=& \frac{1}{4}
\Big( ({\bf h} + |{\bf h}|) \rho_\Sigma{\cal S}C u_1 \Big| 
|{\bf h}|^{-2} ({\bf h} + |{\bf h}|) 
\rho_\Sigma{\cal S}u_2\Big) \nonumber \\
&=& \frac{1}{4}
\Big( \rho_\Sigma (it^a \nabla_a + |{\bf h}|){\cal S} C u_1 \Big| 
|{\bf h}|^{-2} \rho_\Sigma (it^a \nabla_a + |{\bf h}|){\cal S}u_2 \Big),   
\end{eqnarray}
where in the third line we 
have used that $it^a\nabla_a {\cal S}u = 
{\bf h} {\cal S}u$ for all smooth spinor fields $u$ on $M$, by 
the Dirac equation. The key observation is that in an ultrastatic spacetime, we have that 
\begin{equation}
(it^a \nabla_a - |{\bf h}|)(it^b \nabla_b + |{\bf h}|)u = (\nabla^a \nabla_a - \frac{1}{4}R - m^2)u 
\end{equation}
where $R$ is the Ricci scalar. The operator ${\bf h}^2$ is the self adjoint extension on 
$L^2(DM \restriction \Sigma)$ of the elliptic differential operator $q^{ab}\nabla_a \nabla_b + m^2$, 
where $q_{ab}$ is the induced (negative definite) Riemannian metric on $\Sigma$. Therefore, 
by standard results about 
powers of positive elliptic pseudo differential operators on compact manifolds (see e.g. \cite{gil}), 
$|{\bf h}|$ is a pseudo differential operator of order 1 on $\Sigma$ with principal 
symbol $\sqrt{-q^{ab}(x) \xi_a \xi_b} \, I$, where $I$ is the identity map in the fibers of $DM$, 
and $|{\bf h}|^{-2}$ is a pseudo differential operator of order $-2$. 
Equation~\eqref{possib} therefore provides a presentation of the two-point function
${\cal S}_\omega$ to which Thm.~3.12 of~\cite{ju} is applicable, and we conclude by that 
theorem that ${\rm WF}({\cal S}_\omega) \subset C^+$, thus showing that  
${\cal S}_\omega$ is of Hadamard form. The above arguments do not apply as stated 
to the case of non-compact $\Sigma$, since it is not guaranteed then that $|{\bf h}|$
is a pseudo differential operator. This can presumably be shown 
provided that the spatial metric $q_{ab}$ satisfies suitable fall-off conditions at infinity, but 
would require a considerable effort.  

Statement (ii) follows immediately from (i), together with the second remark following Def.~\ref{hadadef}. 

\smallskip

It is known that if the 
split property holds for a state $\omega$, then it also holds for 
any other state that is locally quasiequivalent to $\omega$. Therefore, 
since the ground state on an ultrastatic spacetime is Hadamard by item (i) of the above lemma, and 
since all Hadamard states are locally quasiequivalent by Thm.~\ref{lqe}, 
we conclude that the split-property holds for {\it any} quasifree Hadamard state on 
an ultrastatic spacetime. More generally, by the same argument and (ii) of the above 
lemma, it follows that if a spacetime has a Cauchy surface $\Sigma$ with an 
ultrastatic neighborhood, then the split property holds for any quasifree 
Hadamard state and any pair of concentric double cones $\cO_1$ and $\cO_2$ 
whose bases are in $\Sigma$. 

In order to further generalize this to arbitrary quasifree Hadamard
states on an arbitrary globally hyperbolic
spacetime, we now employ a deformation argument
identical to the one given 
in \cite[Proof of Prop. 5]{v}. For this, one notes that one can construct, 
besides the original, arbitrary globally hyperbolic spacetime $(M, g_{ab})$ (with spin structure), 
a deformed spacetime $(\widehat M, \widehat g_{ab})$ with spin structure, with the 
following properties: 
\begin{enumerate}
\item
$(M, g_{ab})$ possesses a neighborhood $U$ of some Cauchy surface $\Sigma$ which is 
isometric to some neighborhood $\widehat U \subset \widehat M$ containing a Cauchy surface  
$\widehat \Sigma$ in the deformed spacetime $(\widehat M, \widehat g_{ab})$. We chose the spin structure of 
$\widehat M$ so that its restriction to $\widehat U$ 
coincides (via the isometry) with the restriction of spin structure of $M$ to $U$\footnote{It 
then follows that the spin structure over $\widehat M$ can be globally identified with 
the spin structure over $M$, since $M$ resp. $\widehat M$ are topologically 
${\mathbb R} \times \Sigma$ resp. ${\mathbb R} \times \widehat \Sigma$
and since the spin structures over $\Sigma$ resp. $\widehat \Sigma$
have already been identified.}.
\item 
The spacetime $(\widehat M, \widehat g_{ab})$ is ultrastatic in a neighborhood 
of a Cauchy surface $\widehat S$. 
\item
If $\cC_1, \cC_2 \subset \Sigma$ are sets as in the hypothesis of the 
theorem, then there exist open subsets $\widehat \cV_1, \widehat \cV_2 \subset 
\widehat S$ with compact closure such that ${\rm clo}(\widehat \cV_1) \subset 
\widehat \cV_2$ and 
$$\widehat \cC_1 \subset \widehat \Sigma \cap D(\widehat \cV_1), \quad
\widehat \cC_2 \supset \widehat \Sigma \cap J(\widehat \cV_2),$$ where the sets 
$\widehat \cC_i$ correspond to $\cC_i$ and $\widehat \Sigma$ corresponds to $\Sigma$ 
via the isometry postulated in 1).
\end{enumerate}
Now let $\omega$ be an arbitrary quasifree Hadamard state on $\fF$ for the original spacetime, 
and let $\widehat \omega$ be the uniquely determined quasifree Hadamard state for 
the deformed spacetime, whose two-point function coincides with that 
of $\omega$ in $U \times U$ via the isometry identifying the neighborhoods $U$ and
$\widehat U$ described in item 1). The trick 
is now to use the split-property for the deformed spacetime, 
\begin{equation}
\label{defspt}
\pi_{\widehat \omega}(\widehat \fF({\rm int}(D(\widehat \cV_1))))'' \subset \widehat \cN \subset 
\pi_{\widehat \omega}(\widehat \fF({\rm int}(D(\widehat \cV_2))))''.
\end{equation}     
(which is already known, by item 1) and 2) above and by what we have said so far) to prove the split-property 
in the undeformed spacetime. By item 3) above, we have that 
$$
\widehat \cO_1 = {\rm int}(D(\widehat \cC_1)) \subset {\rm int}(D(\widehat \cV_1)), \quad  
\widehat \cO_2 = {\rm int}(D(\widehat \cC_2)) \supset {\rm int}(D(\widehat \cV_2)), 
$$
therefore we have by \eqref{defspt} that
$$
\pi_{\widehat \omega}(\widehat \fF(\widehat \cO_1))'' \subset \widehat \cN \subset 
\pi_{\widehat \omega}(\widehat \fF(\widehat \cO_2))''. 
$$      
Since $\widehat \omega$ agrees with $\omega$ on $U \times U$, this therefore 
shows the split-property on the undeformed spacetime. 
\end{proof}

\vspace{1cm}
\noindent
{\bf Acknowledgements:} 
This work was supported in part by MURST, INDAM-GNAMPA 
and NFS grant PHY00-90138 to the University of Chicago. 
The authors would like to thank R. Brunetti, K. Fredenhagen and E. Schroe
for the invitation to the Oberwohlfach meeting ``Microlocal Analyisis in 
Quantum Field Theory'', 
at which part of this work was completed. 
S. Hollands would like to thank for the kind hospitality and
financial support during his stay at the University of Rome II 
``Tor Vergata'' from January--June 2000.
He would also like to thank R. Conti and 
R. M. Wald for helpful discussions. C.D'Antoni thanks warmly L.Zsido 
and J.E.Robert for helpful discussions. We are in particular indebted to the 
referee, who pointed out an important shortcoming in our discussion
of quasiequivalence in the original version of this paper. 

\appendix

\section{Appendix}

\subsection{Proof of (iii) of Lem.~\ref{lem1}}

We here sketch how to obtain the estimate $\|S\|_1 = \|T\|_1 \le (\beta_0/\beta)^s e^{-\beta m_0/2}$ claimed
in item (iii) of Lem. \ref{lem1}. The method of our proof also allows us to establish that the operators 
${\bf k}_n$ introduced in the proof of Lem. \ref{lem1} are of Hilbert-Schmidt class, as had been 
claimed there. However for brevity, will explicitly
demonstrate this only for the case $n = 1$, the other cases can be treated in a similar way.

In order to prove the above estimate on the trace-norm of $S$, 
we write $S = \sqrt{2} E_\cC \cdot {\bf k}_1 \cdot {\bf k}_2 \cdot
(\myid + \beta^2 {\bf h}^2)^{s} e^{-\beta {\bf h}} P$ 
as in the proof of Lem. \ref{lem1}, where 
$$
{\bf k}_1 = M_\chi (\myid + \beta^2 {\bf h}^2)^{-s/2}, \quad 
{\bf k}_2 = (\myid + \beta^2 {\bf h}^2)^{s/2} M_\chi 
(\myid + \beta^2 {\bf h}^2)^{-s},  
$$  
and where $M_\chi k = \chi k$ is the multiplication operator by a 
compactly supported smooth function 
$\chi$ which is identically one on $\cC$. We get from this the estimate
$$
\| S \|_1 \le \| {\bf k}_1 \|_2 \| {\bf k}_2 \|_2  \|(\myid + \beta^2{\bf h}^2)^{s} e^{-\beta {\bf h}} P\|
\le c \| {\bf k}_1 \|_2 \| {\bf k}_2 \|_2  e^{-\beta m_0/2}.
$$
Hence, in order to prove the estimate in item (iii) of Lem. \ref{lem1}, it is sufficient 
to demonstrate that $\|{\bf k}_1\|_2 \le c\beta^{-s/2}$ and that $\|{\bf k}_2\|_2 \le c\beta^{-s/2}$, 
where $c$ is some constant independent of 
$\beta$. We will now show how to obtain the bound on the Hilbert-Schmidt norm of  ${\bf k}_1$. 
The bound on the Hilbert-Schmidt norm of ${\bf k}_2$, and likewise 
on the other ${\bf k}_n$, can be established in a similar way, 
but we shall not discuss this here. 

The task is thus to prove that 
\begin{equation}
\label{bound}
\|{\bf k}_1 \|_2^2 = \| M_\chi (\myid + \beta^2 {\bf h}^2)^{-s} M_\chi \|_1 \le c\beta^{-s}.
\end{equation}
Let us assume, for simplicity, that $\supp(\chi) \subset \Sigma$ can be covered by a single 
chart $X \subset \mr^s$, and identify $\cC$  
via this chart with a subset of $X$.\footnote{If $\cC$ cannot be covered by single coordinate 
chart, one can write $\chi = \sum \chi_i$, where each $\chi_i$ is supported in a 
single chart. The argument then still goes through but we do not give any details here.} 
The idea of the  
proof is to split $M_\chi (\myid + \beta^2{\bf h}^2)^{-s} M_\chi$, viewed 
now as a self-adjoint 
operator on $L^2(X; \mc^N)$ (where $N = 2^{[(s+1)/2]}$ is the number of spinor components
in $s+1$ dimensions), into two parts which are more amenable to an analysis than this 
operator itself. In order to achieve this, we must first recall the notion 
of a pseudo differential operator with a parameter (for details, see e.g. 
\cite[p.58]{gil}). These are defined---just as ordinary pseudo differential
operators---in terms of so-called symbols, the only difference being that these symbols 
now depend on an additional parameter, $\lambda$, which we shall take to be 
$\beta^{-2}$ later on. Let $S^m_\lambda(X)$ be the space of all matrix valued functions 
$p(x, \xi, \lambda)$ which are smooth in $x \in X$ and $\xi \in \mr^s$ and 
analytic in $\lambda \in \mc\backslash\mr^-$ and for which 
$|\partial^\alpha_\xi \partial^\gamma_x p | \le c_{\alpha,\gamma} 
(1 + |\xi| + |\lambda|^{1/2})^{m - 
|\alpha|}$, where we use the usual multiindex notation, for example 
$\partial^\alpha_x = 
\frac{\partial^{\alpha_1}}{\partial x_1^{\alpha_1}} \dots \frac{\partial^{\alpha_s}}{\partial x_s^{\alpha_s}}.$ 
Elements in $S^m_\lambda(X)$ are
called ``symbols with a parameter''. A pseudo differential operator
with parameter $\lambda$ is an operator of the form 
$$P_\lambda u(x) = (2\pi)^{-s/2}\int_{\mr^s} p(x, \xi, \lambda) e^{ix\xi} \widehat{u}(\xi) \, d^s \xi, $$
where $p$ is a symbol with parameter and $\widehat{u}$ is the Fourier transform of $u$. 
It is customary to write $\sigma P_\lambda = p$. The space of all pseudo 
differential operators $P_\lambda$ with $\sigma P_\lambda \in S^m_\lambda(X)$ is
denoted by $\Psi^m_\lambda(X)$.  

It follows form \eqref{hdef} that 
${\bf h}^2$ is a self adjoint extension of an elliptic differential operator 
$A$ with symbol 
\begin{equation}
\sigma A(x, \xi) = a_2(x, \xi) + a_1(x, \xi) + a_0(x, \xi).    
\end{equation}
The principal symbol $a_2(x, \xi)$ is given by the matrix 
$q^{ij}(x) \xi_i \xi_j I$ in a local coordinate chart, 
where $I$ denotes the $N \times N$ identity matrix and 
$q^{ij}$ are the coordinate components of the inverse 
spatial metric. $a_1$ and $a_0$ are symbols of order 1 
and 0, respectively, whose particular form is not 
relevant for our purposes. 

We define the symbols
\begin{eqnarray*}
b_0 &=& (a_2 + \lambda I)^{-1} \quad \in S^{-2}_\lambda(X),\\
b_k &=& -(a_2 + \lambda I)^{-1} \sum_{2 + |\alpha| + j - i = k, j<k} i^{|\alpha|}
\partial^\alpha_\xi b_j \partial^\alpha_x a_i/\alpha! \quad \in S^{-2-k}_\lambda(X). 
\end{eqnarray*}
Let $B_\lambda \in \Psi^{-2}_\lambda(X)$ be a pseudo differential operator  whose symbol 
has the asymptotic expansion
\begin{equation}
\sigma B_\lambda \sim \sum_{k \ge 0} b_k(x, \xi, \lambda) 
\quad \in S^{-2}_\lambda(X). 
\end{equation}
By construction, $B_\lambda$ is an inverse modulo $\Psi^{-\infty}_\lambda(X)$
of $A$ on $X$ (see e.g. \cite{gil}), in the sense that  
\begin{equation}
B_\lambda (\lambda \myid + A)  = (\lambda \myid + A) B_\lambda = \myid \quad 
\text{modulo $\Psi^{-\infty}_\lambda(X)$.}  
\end{equation}
Moreover, since $A$ is formally self adjoint and positive, 
it follows from well known arguments \cite{gil}
that the operator $B_\lambda$ can be assumed to be a positive operator (this can 
be achieved, if necessary, by adding a suitable element in $\Psi^{-\infty}_\lambda(X)$
to $B_\lambda$). 

We now write
\begin{equation}
M_\chi (\lambda\myid + {\bf h}^2)^{-s} M_\chi = Q_\lambda  + R_\lambda,
\end{equation}
where $\lambda > 0$ from now on and 
\begin{equation}
Q_\lambda = M_\chi B_\lambda^s M_\chi, \quad
R_\lambda = M_\chi ((\lambda \myid + A)^{-s} - B_\lambda^s) M_\chi. 
\end{equation} 
(Here and in the following we write the operator $(\lambda \myid + {\bf h}^2)^{-s}$
as $(\lambda \myid + A)^{-s}$, to simplify the notation. Note that the latter is 
is understood as an operator on $\cK$. This operator is not assumed to be 
pseudo differential, nor must it be confused with $B_\lambda^s$, which is pseudo
differential, but only defined on $X$.)

The relevant properties of $R_\lambda, Q_\lambda$ are summarized in the following 
lemma:
\begin{lem}\label{l4}
We have $\|Q_\lambda\|_1 \le c\lambda^{-s/2}$ and $\|R_\lambda\|_1 \le c\lambda^{-s/2}$
for some constant $c$.  
\end{lem}

The lemma allows us to finish the proof of eq.~\eqref{bound}
because, setting $\lambda = \beta^{-2}$,  
\begin{eqnarray*}
\|M_\chi(\myid + \beta^2 {\bf h}^2)^{-s} M_\chi \|_1 \le \beta^{-2s} (
\|Q_{\beta^{-2}}\|_1 + \|R_{\beta^{-2}}\|_1) \le \beta^{-2s} \cdot c\beta^s = c\beta^{-s}.
\end{eqnarray*}

\qed

It remains to prove lemma~\eqref{l4}.

\begin{proof}
Since $Q_\lambda$ is a positive operator in $\Psi^{-2s}_\lambda(X)$, 
we can calculate its trace norm by (here and in the following 
we use the constant convention, meaning that 
different numerical constants, are denoted by the same symbol $c$):
\begin{eqnarray*}
\|Q_\lambda\|_1 = \tr_{L^2(X; \mc^{N})} Q_\lambda &=& 
(2\pi)^{-s/2} \tr_{\mc^{N}}
\int_X \int_{\mr^s} q(x, \xi, \lambda) \,d^s\xi d^s x \\
&\le& c \int_{\mr^s} (1 + |\xi| + \lambda^{1/2})^{-2s} \,d^s\xi\\
&\le& c \int_0^\infty (1 + r + \lambda^{1/2})^{-2s} r^{s-1} \,dr\\
&=& c(1+ \lambda^{1/2})^{-s} \le c\lambda^{-s/2}. 
\end{eqnarray*}
This proves the first inequality. For notational simplicity, we will only
explicitly prove the second inequality for the case $s=1$, but our arguments
can be generalized straightforwardly to deal with general values for $s$.
We first introduce some notation. Let $\chi', \chi'', \chi''', \dots$ be smooth 
functions of compact support in $X$ such that $\chi' \equiv 1$ on the support 
of $\chi$, $\chi'' \equiv 1$ on the support of $\chi'$, $\chi''' \equiv 1$ on 
the support of $\chi''$, etc. Furthermore, let us set 
\begin{equation}
L = [M_\chi, A], \quad L' = [M_{\chi'}, A], \quad L'' = [M_{\chi''}, A], \dots,  
\end{equation}
so that $L, L', L'', \dots$ are partial differential operators of degree 1.
Let $\|\, \cdot \,\|_{H^r}$ denote the norm on the Sobolev space
$H^{r}(X; \mc^{N})$, and let $k$ be an arbitrary smooth spinor field. 
Since $A$ is elliptic, we obtain the estimate:
\begin{eqnarray*}
\|R_\lambda k\|_{H^r} &=&
\| M_\chi ((\lambda \myid + A)^{-1} - B_\lambda)M_\chi k \|_{H^r}\\
&\le& c \|(\lambda\myid + A) M_\chi ((\lambda \myid + A)^{-1} - B_\lambda)
M_\chi k \|_{H^{r-2}}
\end{eqnarray*}
If we move the operator $(\lambda \myid + A)$ through $M_\chi$, then 
this can be further estimated by 
\begin{eqnarray*}
&\le& 
c \Big( 
\|M_\chi(\myid - (\lambda \myid - A)B_\lambda)M_\chi k \|_{H^{r-2}} + 
\|LM_{\chi'}((\lambda \myid - A)^{-1} - B_\lambda)M_\chi k \|_{H^{r-2}}  
\Big)\\
&\le& 
c \Big( 
(1 + \lambda)^{-m/2} \| k \|_{L^2} + 
\|M_{\chi'}((\lambda \myid - A)^{-1} - B_\lambda)M_\chi k \|_{H^{r-1}}  
\Big),   
\end{eqnarray*}
where we have used that $\myid - (\lambda \myid + A)B_\lambda  
\in \Psi^{-\infty}_\lambda(X)$
in the last line, and where $m$ is an arbitrary but fixed natural number. 
We may repeat the above chain of inequalities for the second expression in the 
last line, replacing $\chi'$ by $\chi''$ and $L$ by $L'$. If we iterate this procedure, 
we get
\begin{eqnarray*}
\|R_\lambda k\|_{H^r} 
&\le& 
c \Big( 
(1 + \lambda)^{-m/2} \| k \|_{L^2} + 
\|M_{\chi''}((\lambda \myid - A)^{-1} - B_\lambda)M_\chi k \|_{H^{r-2}}  
\Big)\\
&\le& 
c \Big( 
(1 + \lambda)^{-m/2} \| k \|_{L^2} + 
\|M_{\chi'''}((\lambda \myid - A)^{-1} - B_\lambda)M_\chi k \|_{H^{r-3}}  
\Big)\\
&\le& \dots\\
&\le& 
c \Big( 
(1 + \lambda)^{-m/2} \| k \|_{L^2} + 
\|M_{\chi^{\prime \prime \dots \prime}}((\lambda \myid - A)^{-1} - B_\lambda)M_\chi k \|_{L^2}  
\Big)\\
&\le& 
c (1 + \lambda)^{-m/2} \| k \|_{L^2},   
\end{eqnarray*}
where neither of the above constants depend on $\lambda$, but only on $r$ and $m$. This inequality 
implies that $R_\lambda$ is a continuous map from $L^2(X;\mc^N)$ to $H^r(X;\mc^N)$ for
every $r$. By duality, $R^*_\lambda = R_\lambda$ is a continuous map from 
$H^{-r}(X;\mc^N)$ to $L^2(X;\mc^N)$ for every $r$. Therefore, $K_\lambda = R_\lambda R_\lambda^*$
is a continuous map $K_\lambda: H^{-r}(X;\mc^N) \to H^{r}(X;\mc^N)$ for every $r$ and therefore 
has a smooth (and by definition compactly supported) integral kernel on $X \times X$. 
From the dependence upon $\lambda$ of the constants in the above inequality, one 
gets furthermore that 
\begin{eqnarray}
\label{**}
\sup |\partial^\alpha_x \partial^\gamma_y K_\lambda(x, y) | \le 
c_{\alpha, \gamma}
(1 + \lambda)^{-m} \quad \forall \alpha,\gamma.
\end{eqnarray}
This estimate can now be used to bound the $p$-th Schatten norm $\| \, \cdot \, \|_p$ of 
$K_\lambda$ by writing $K_\lambda$ as a product of an arbitrary number of Hilbert-Schmidt
operators as in eq.~\eqref{lwrite}, 
\ben
K_\lambda(x,y) = \prod_i^n (1+ |x|)^{-q} \cdot (1+ |x|)^{nq} K_\lambda(x,y), 
\een
where the $q$ is chosen sufficiently large. The multiplication operator by $(1+|x|)^{-q}$ is then Hilbert-Schmidt, 
and $\|(1+ |x|)^{nq} K_\lambda\|_2 \le c(1 + \lambda)^{-m}$ using eq.~\eqref{**}. Hence
\ben
\label{kest}
\|K_\lambda\|_p \le c(1 + \lambda)^{-m}
\een
for all $p > 0$ and $\lambda > 0$, where $c$ depends only on $p,m$. The chain of inequalities leading 
to eq.~\eqref{kest} can be repeated for arbitrary $s$. 
It follows in particular 
that $\|R_\lambda\|_1^2 = \| K_\lambda \|_{1/2} \le c(1 + \lambda)^{-m} \le c\lambda^{-s}$ (choosing 
$m = s$). This the claim of the lemma for $R_\lambda$. 
\end{proof}

\subsection{Nuclear maps}
\label{sec2.2}
The notion of nuclearity considered in this paper is formulated in 
terms of nuclear maps. For the convenience of the reader, 
we now recall the definition of a nuclear map between two Banach spaces, 
for details we refer to \cite{p}.  

Let $\cX$ and $\cY$ be Banach spaces, with norms $\|\, \cdot \,\|_\cX$ and
$\|\, \cdot \,\|_\cY$ respectively. A bounded linear map $\Theta: \cX
\to \cY$ is called {\it nuclear}, if there exist bounded linear functionals 
$\varphi_i$ on $\cX$ and vectors $Y_i \in \cY, i \in \mi$ 
such that $\Theta$ can be written as 
\begin{equation}
\label{twrite}
\Theta(X) = \sum_i Y_i \varphi_i(X) \quad \forall X \in \cX,   
\end{equation}
with
$$ \sum_i \|\varphi_i\| \, \|Y_i\|_\cY < \infty, 
\quad \|\varphi_i\| = \sup_{\|X\|_\cX = 1} |\varphi_i(X)|.$$ 
One defines the {\it nuclearity index} of $\Theta$ by 
$$
\|\Theta\|_1 = \inf \sum_i \|\varphi_i\| \, \|Y_i\|_\cY, 
$$
where the infimum is taken over all possible ways to write 
$\Theta$ in the form \eqref{twrite}. The set of all nuclear maps is
denoted by $\cI_1(\cX, \cY)$. More generally, a bounded linear map
$\Theta$ is called {\it p-nuclear}, if it can be written in the 
form \eqref{twrite} with 
$$\|\Theta\|_p = \inf \left( \sum_i \|\varphi_i\|^p \, \|Y_i\|_\cY^p \right)^{1/p}
< \infty, \quad p > 0.$$
The space of all $p$-nuclear maps, $p > 0$, from $\cX$ to $\cY$ is denoted 
by $\cI_p(\cX, \cY)$. It can be shown that these spaces are again
Banach spaces for all $p > 0$. If $\cX = \cY$, then we simply write
them as $\cI_p(\cX)$. If $\cX$ is a Hilbert space, then $\cI_1(\cX)$ is
the space of trace-class operators, and $\cI_2(\cX)$ is 
the space of Hilbert-Schmidt operators.


\end{document}